\documentclass[prm,reprint,amsmath,amssymb,linenumbers]{revtex4-2}
\usepackage{graphicx}
\usepackage{natbib}
\usepackage{bm}
\usepackage{float}
\usepackage{array}
\usepackage[usenames]{color}

\begin{document}

\title{Piezoelectric Electrostatic Superlattices in Monolayer MoS$_2$}

\author{Ashwin Ramasubramaniam}
\email{ashwin@umass.edu}
\affiliation{Department of Mechanical and Industrial Engineering, University of Massachusetts Amherst, Amherst MA 01003}
\affiliation{Materials Science and Engineering Graduate Program, University of Massachusetts Amherst, Amherst, MA 01003}

\author{Doron Naveh}
\email{doron.naveh@biu.ac.il}
\affiliation{Faculty of Engineering, Bar-Ilan University, Ramat Gan 52900, Israel}
\affiliation{Institute for Nanotechnology and Advanced Materials, Bar-Ilan University, Ramat Gan 52900, Israel}

\date{\today}

\def\linenumberfont{\normalfont\tiny\sffamily}

\begin{abstract}
Modulation of electronic properties of materials by electric fields is central to the operation of modern semiconductor devices, providing access to complex electronic behaviors and greater freedom in tuning the energy bands of materials. Here, we explore one-dimensional superlattices induced by a confining electrostatic potential in monolayer MoS$_2$, a prototypical two-dimensional semiconductor. Using first-principles calculations, we show that periodic potentials applied to monolayer MoS$_2$ induce electrostatic superlattices in which the response is dominated by structural distortions relative to purely electronic effects. These structural distortions reduce the intrinsic band gap of the monolayer substantially while also polarizing the monolayer through piezoelectric coupling, resulting in spatial separation of charge carriers as well as Stark shifts that produce dispersive minibands. Importantly, these minibands inherit the valley-selective magnetic properties of monolayer MoS$_2$, enabling fine control over spin-valley coupling in MoS$_2$ and similar transition-metal dichalcogenides.
\end{abstract}

\nolinenumbers
\maketitle


\section{Introduction}

The electronic band structure of a solid—a representation of allowed electronic energy levels—reflects the quantum mechanical response of electrons to the underlying periodic potential of the atomic lattice. A superlattice induces further periodic modulations of the lattice potential over length scales several times larger than the atomic spacing yet smaller than the mean free path of an electron, localizing free carriers and causing further quantization of energy bands into minibands. \cite{esaki_superlattice_1970,smith_theory_1990} While semiconductor superlattices are typically produced by alternating layers of different semiconductors or by differently doped (n-/p-type) layers of a single semiconductor, it is also possible to induce directly a superlattice potential via externally defined electrostatic gates or patterned dielectric substrates.\cite{forsythe_band_2018,li_anisotropic_2021,huber_gate-tunable_2020,barcons_ruiz_engineering_2022,dragoman_bloch_2021,kyoungryu_superlattices_2019} Such externally-defined electrostatic superlattices (henceforth, simply “electrostatic superlatttices”), employed early on with two-dimensional (2D) electron gases, \cite{schmeller_franzkeldysh_1994,schlosser_landau_1996,zimmermann_lateral_1997}  continue to attract interest in 2D materials, \cite{forsythe_band_2018,li_anisotropic_2021,huber_gate-tunable_2020,barcons_ruiz_engineering_2022,dragoman_bloch_2021,kyoungryu_superlattices_2019,yankowitz_van_2019} motivated in part by theoretical studies of 2D superlattices that promise key advances, for example, in supercollimation, Mott-insulating states, superconductivity, and topological subbands.\cite{choi_electron_2014,balents_superconductivity_2020,behura_moire_2021,chen_evidence_2019,cazalilla_quantum_2014,ghorashi_topological_2023} It is worth noting that superlattices in 2D materials can also be achieved by various other means, for example, by exploiting the naturally occurring periodic potential of Moir{\'e} patterns \cite{he_moire_2021,kennes_moire_2021,kogl_moire_2023,dean_hofstadters_2013} or via strain patterning, \cite{li_optoelectronic_2015,banerjee_strain_2020,zhang_magnetotransport_2019} among others. An assessment of these various approaches can be found in a recent review.\cite{kyoungryu_superlattices_2019}

To date, studies of 2D electrostatic superlattices—either experimental \cite{forsythe_band_2018,li_anisotropic_2021,huber_gate-tunable_2020,barcons_ruiz_engineering_2022,dragoman_bloch_2021,kyoungryu_superlattices_2019,yankowitz_van_2019}  or theoretical \cite{choi_electron_2014,park_electron_2008,lin_periodically_2020,hui-yun_tunable_2011,chen_electrostatic_2020,brey_emerging_2009,wu_graphene_2012}—have focused primarily on graphene (a semi-metal). In comparison, electrostatic superlattices in 2D transition-metal dichalcogenides (TMDCs) are sparsely studied, that too only at the theoretical level \cite{ono_effect_2017,sattari_effect_2021} without accounting for the role of \emph{potential-induced} strains. However, not only is it well known that the (opto)electronic properties of TMDCs are sensitive to strain \cite{chaves_bandgap_2020,conley_bandgap_2013,manzeli_piezoresistivity_2015,yun_thickness_2012} but it is also well known that inhomogeneous strains can lead to spatial variations in band profiles that cause, for example, funneling of excitons and trions. \cite{li_optoelectronic_2015,feng_strain-engineered_2012,lee_switchable_2021,lee_drift-dominant_2022} In contrast, this coupling between strain and potential has been studied widely in Moir{\'e} superlattices of TMDCs (homo-/hetero-bilayers) and it is recognized that accounting for spatial variations of strain and morphology is essential for modeling the response of these structures correctly.\cite{waters_flat_2020,rodriguez_complex_2023,quan_phonon_2021,zhu_moire-templated_2018,zhang_interlayer_2017}  Thus, in this paper, we take a prototypical example of a one-dimensional (1D) electronic superlattice in monolayer MoS$_2$ and seek to understand how coupling between the external electrostatic potential and induced strains affects the electronic response of the material. We employ first-principles density functional theory calculations and show that significant atomic relaxation is possible within the MoS$_2$ monolayer, when subjected to a periodic external potential, and that these effects can lead to over an order of magnitude decrease in the superlattice band gap than would be anticipated from a purely electronic response. Moreover, since monolayer MoS$_2$ is piezoelectric,\cite{duerloo_intrinsic_2012} we find a strong anisotropic, spatially-varying coupling of the polarization response to the strain field induced by the external superlattice potential, and we show that the internal polarization fields significantly influence mini-band formation. Importantly, we find that the mini-bands formed at the band edges preserve the appealing spin–valley coupling property of monolayer MoS$_2$, opening up possibilities for spin-selective mini-band engineering of MoS$_2$ and analogous 2D semiconductors.

\section{Results and Discussion}
\begin{figure}[htbp!]
\centering
\includegraphics[width=\linewidth]{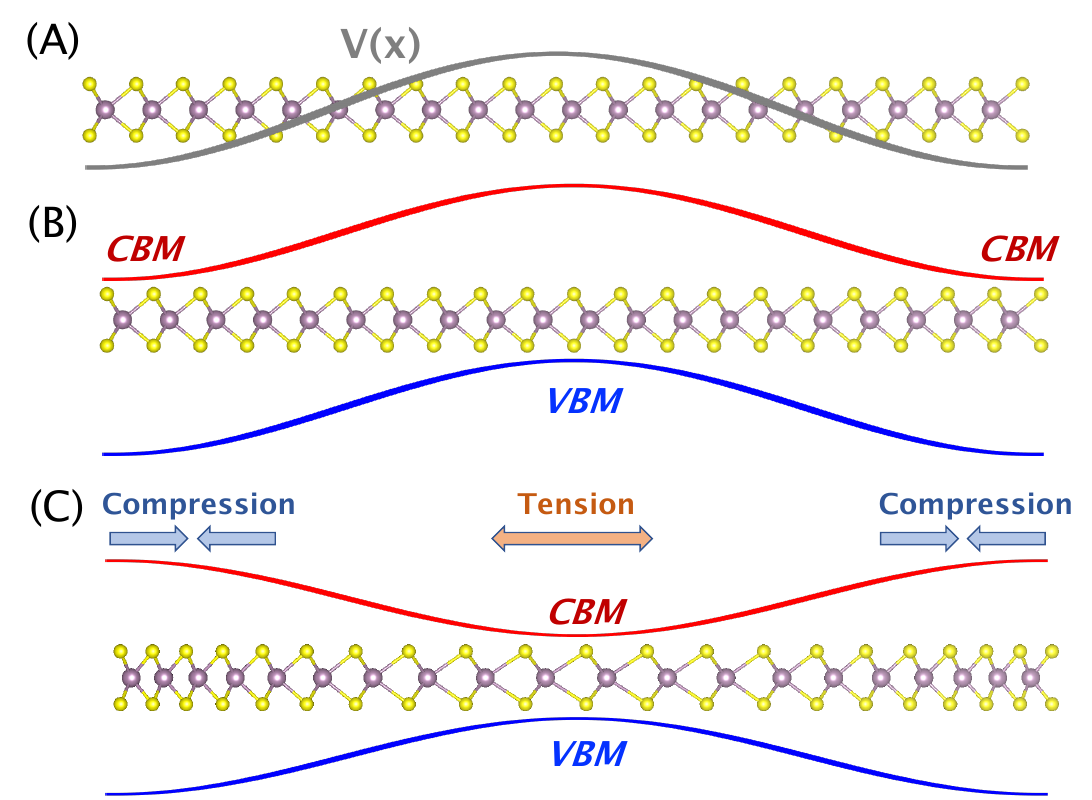}
\caption{\label{fig1} (A) A 2D material (e.g., monolayer MoS$_2$) subjected to a spatially-periodic external potential $V(x)$. (B) In the absence of structural relaxation, shifts in the electronic energy levels follow the potential modulation, leading to localization of the valence band maximum (VBM) and conduction band minimum (CBM) in the regions of highest and lowest potential, respectively. (C) Structural relaxation leads to displacement of atoms away from regions of high potential and towards regions of low potential, inducing tensile and compressive strains, respectively. Tensile strain reduces the band gap of the monolayer while compressive strain increases the band gap and the VBM and CBM are now localized within the region of higher potential.
}
\end{figure}

The key concept of this paper is illustrated schematically in Fig.\ref{fig1}: we seek to understand the response of a 2D semiconductor (e.g., monolayer MoS$_2$) subjected to a spatially-periodic potential, $V(x)$, whose magnitude is much smaller than the potential of the nuclei. If we ignore the forces induced by the applied potential on the atoms, the response of the material is purely electronic, and the energy levels rise and fall in phase with the potential modulation (Fig.\ref{fig1}B). As a consequence, the valence band maximum (VBM) and the conduction band minimum (CBM) are localized at the corresponding maximum and minimum of the applied potential and are spatially separated. In analogy with the more familiar situation of semiconductor heterojunctions, the material response is akin to a staggered-gap (type II) heterojunction. \cite{schmeller_franzkeldysh_1994}  However, if structural relaxation is accounted for, as it should be, atoms are displaced from regions of higher potential energy towards regions of lower potential energy, leading to a distribution of tensile and compressive strains within the material (Fig.\ref{fig1}C). It is well known that the electronic structures of 2D materials are sensitive to strain\cite{chaves_bandgap_2020}: in the case of monolayer MoS$_2$, the band gap decreases or increases under tension and compression, respectively. \cite{manzeli_piezoresistivity_2015,conley_bandgap_2013} Thus, the shifts in electronic energies are no longer merely in phase with the applied potential and, depending upon the precise balance between strain- and potential-induced effects, the VBM and CBM can instead localize within the same region of high tensile strain where the band gap is smallest. This scenario is now akin to a straddling-gap (type I) heterojunction. It is precisely this coupling between external fields and induced strains, and the ultimate consequences for the electronic response of the monolayer, that we explore in detail here. Additionally, for a material such as MoS$_2$, the electronic response is much richer due to its intrinsic piezoelectricity \cite{duerloo_intrinsic_2012,wu_piezoelectricity_2014}  and spin–valley coupling, \cite{xiao_coupled_2012,mak_control_2012} which we explore in detail below.

\subsection{Electric field-induced piezoelectric coupling}
\begin{figure*}[htbp!]
\centering
\includegraphics[width=0.8\linewidth]{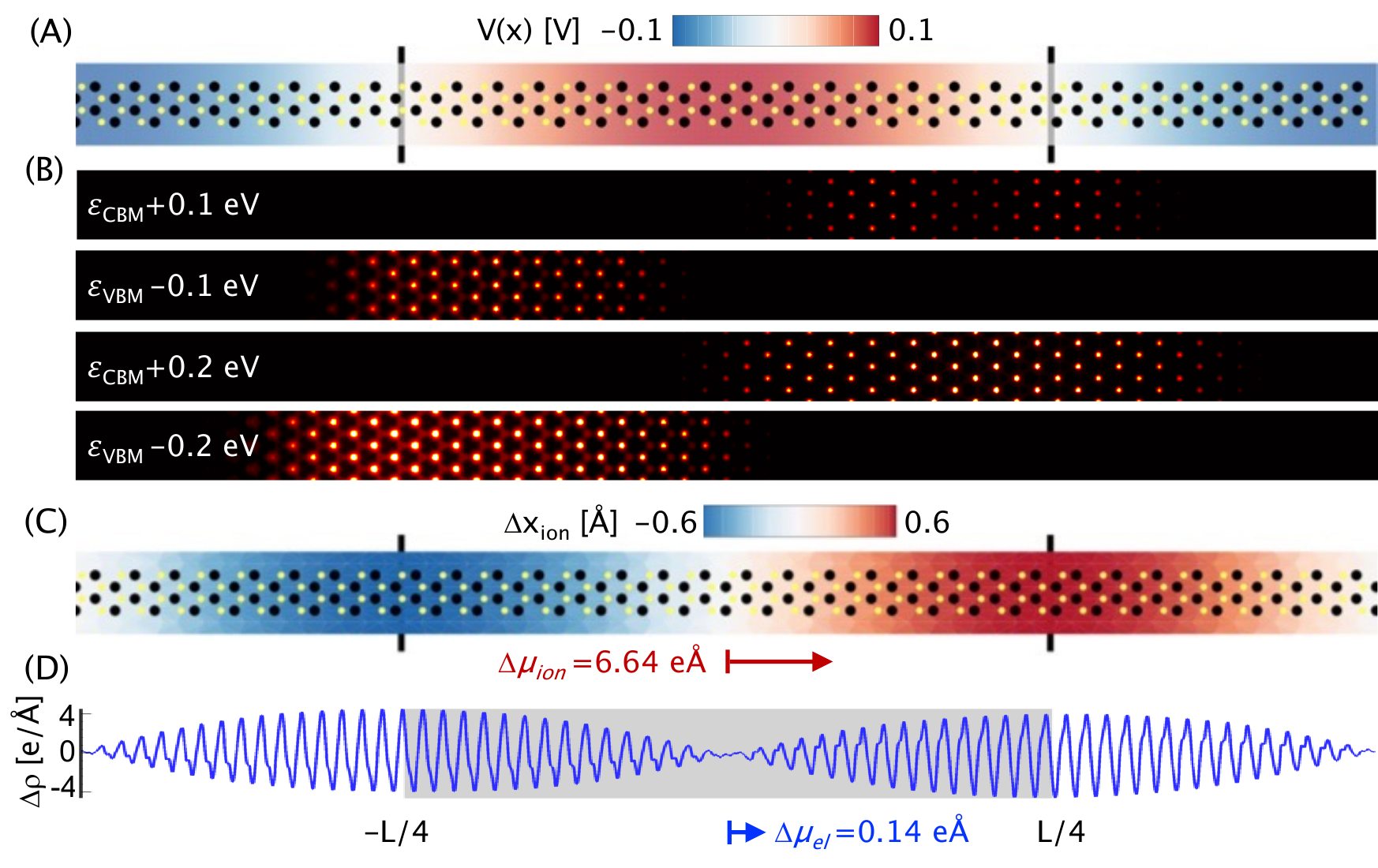}
\caption{\label{fig2} (A) Contour plot of external potential $V(\bm{r})=V_0 \cos(\bm{q}_{AC}\cdot \bm{r})$ with $V_0$=0.1V superimposed over simulation domain consisting of $N_{AC}$=32 repeat units (L=17.6 nm) along the armchair (x-) direction. Thick black lines demarcate the region $-L/4\le x\le L/4$. (B) Integrated electronic charge density arising from windows of $\pm0.1$ eV and $\pm$0.2 eV at the valence (VBM) and conduction (CBM) band edges projected over the plane of the monolayer (x-y plane). (C) Contour plot of atomic displacements induced by the superlattice potential, leading to a net ionic dipole of $\Delta\mu_{ion}=6.64$e\AA within the domain $-L/4\le x\le L/4$. (D) Integrated (over y-z plane) change in electronic charge density $\Delta\rho(x)$ induced by the superlattice potential, leading to a net electronic dipole of $\Delta\mu_{el}=0.14$ eÅ within the domain $-L/4\le x\le L/4$. Black and yellow spheres in (A, C) indicate Mo and S atoms; the simulation cell is repeated along the y-direction in (A-C) for ease of viewing.
}
\end{figure*}

In this work, we restrict attention to one-dimensional potential modulations, $V(\bm{r})=V_0 \cos(\bm{q}\cdot \bm{r})$, with the vector $\bm{q}$ directed along either the armchair $(\bm{q}_{AC})$ or the zigzag $(\bm{q}_{ZZ})$ direction:  Figure \ref{fig2}A displays an example of a superlattice along the armchair direction with $V_0$=0.1 V and period $L$=17.6 nm. The functional form of this potential is motivated by the calculated displacement fields in Ref.\,\onlinecite{forsythe_band_2018} and we assume that the field does not vary appreciably across the thickness of the monolayer. In response to this external perturbation, atoms are displaced from regions of higher potential towards regions of lower potential (peak displacement of $\pm 0.6$\AA; Fig. \ref{fig2}C), leading to tensile and compressive strains, $\epsilon_{11}$, that range between $\pm1.8\%$ (Fig.\,S2). Figure \ref{fig2}B displays the spatial distribution of band edges from which we observe that electronic states, as deep as 0.1-0.2 eV into the valence or conduction band, are predominantly localized within the region of tensile strain. The perturbation reduces the overall band gap by $\sim30\%$ relative to the unperturbed bandgap of monolayer MoS$_2$  (from 1.67 eV to 1.13 eV). By comparison, if atoms are held fixed at their original positions (no structural relaxation; Fig.\,S3), the valence and conduction band edges are spatially separated, localizing within regions of positive and negative potential, respectively, that too becoming more distinct only at larger fields. The decrease in the band gap is now less than 1\% ($\sim15$ meV), over an order of magnitude smaller than the relaxed case at the same applied potential. These observations essentially confirm, at least to lowest order, the simple physical picture presented previously. Of course, the armchair direction is not special in this regard, and a calculation for a cosine potential applied along the zigzag direction shows analogous atomic displacements and localization of band edges within the region of tensile strain, (Fig.\,S4). However, there is one crucial distinction between the responses of the armchair and zigzag directions: for the armchair case, the VBM and CBM, though localized within the region of tensile strain, have very little spatial overlap (Fig.\,\ref{fig2}B); in contrast, for the zigzag case, the VBM and CBM overlap spatially within the region of tensile strain (Fig.\,S4). To understand this markedly anisotropic response, we undertake a more detailed analysis of the structural response in each case.  

Monolayer MoS$_2$ (space group $P\bar{6}m2$; point group $D_{3h}$) lacks a center of inversion and is strongly piezoelectric. \cite{duerloo_intrinsic_2012} The polarization vector ($\bm{P}$) and strain tensor ($\bm{\epsilon}$) are related via the constitutive relation $\bm{P}=\bm{e}{\bm \epsilon}$, where $\bm{e}$ is the third-rank piezoelectric tensor. \cite{nye_physical_1985} When the external potential $V$ is aligned with the armchair direction ($\bm{q}\parallel\bm{q}_{AC}$), the piezoelectric coupling to the induced strain field induces a polarization field that is aligned with the armchair direction ($\bm{P}\parallel\bm{q}_{AC}$). In contrast, when the external potential $V$ is aligned with the zigzag direction ($\bm{q}\parallel\bm{q}_{ZZ}$), the polarization field is along the orthogonal armchair direction ($\bm{P}\perp\bm{q}_{ZZ}$). 
Further details can be found in the supplementary text. 
\footnote{See Supplemental Material at [URL will be inserted by publisher] for details.} 
Considering the armchair case and focusing on the region of tensile strain $-L/4\le x\le L/4$ where the band edges are localized, we calculate a net ionic dipole moment of 6.64 e\AA~or $\sim12 \mu\textrm{C/cm}^2$ along the direction of quantization ($\bm{q}_{AC}$). 
\footnote{The areal dimensions of the supercell are 176.41 \AA $\times$ 3.18 \AA~while the thickness of the monolayer is 3.13 \AA.} 
Note that this degree of polarization is comparable to the remnant polarization in several ferroelectric materials.  \cite{horiuchi_organic_2008} Thus, separation of the valence and conduction band edges within the tensile region is driven by the internal electric field that arises from piezoelectric coupling to the strain field induced by the external potential. For a larger external potential ($V_0$=0.3 V; Fig.\,S5), both the localization of band edges within the region of tensile strain as well as their spatial separation are more marked, as deep as 0.5 eV into the valence/conduction band. By contrast, for the zigzag case, the dipole moments within the region of tensile strain $-L/4\le x\le L/4$ are orthogonal to the direction of quantization and, hence, there is no further spatial separation of band edges within this region of quantum confinement.

\subsection{Minibands and band-gap tuning}

\begin{figure*}[htbp!]
\centering
\includegraphics[width=0.9\linewidth]{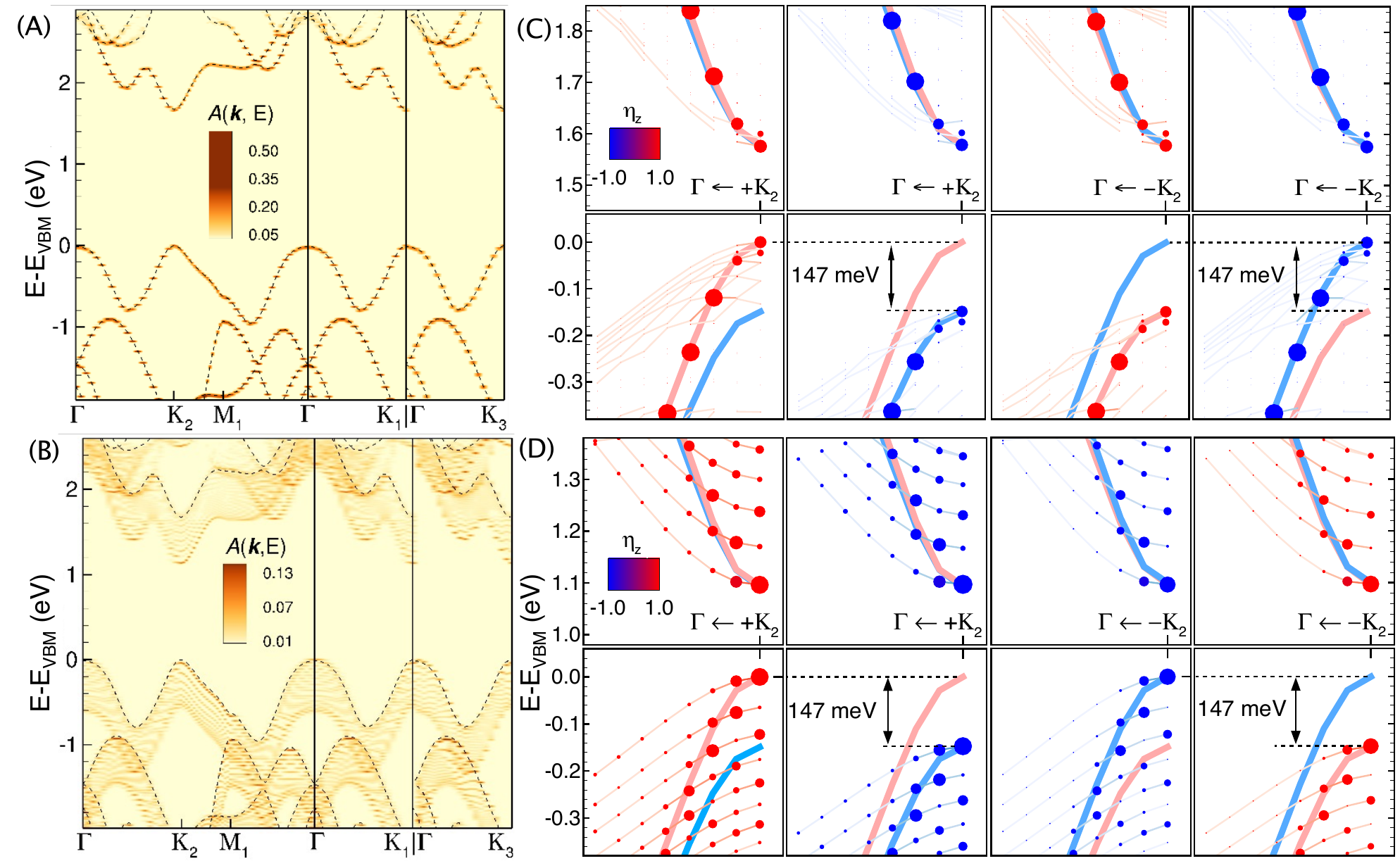}
\caption{\label{fig3} (A, B) Unfolding into the first Brillouin zone of the electronic bandstructure of an MoS$_2$  superlattice ($N_{AC}$=32) subjected to an external potential $V(\bm{r})=V_0\cos(\bm{q}_{AC}\cdot\bm{r})$ with $V_0=0.1$V, without (A) and with (B) atomic relaxation. The intensity of the color map indicates the overall contribution from supercell eigenstates to a primitive cell eigenstate at wavevector $\bm{k}$ and energy E -- the spectral function $A(\bm{k}, E)$.  \cite{dirnberger_electronic_2021} Bands of the pristine (unperturbed) MoS$_2$  monolayer are overlaid in thin dashed lines. (C, D) Details of the unfolded electronic bandstructure at the $\pm K_2$ valleys with the inclusion of spin-orbit splitting without (C) and with (D) atomic relaxation. Symbols are colored by the degree of spin polarization ($\eta_z=m_z/\left|m\right|$); the size of the symbol is proportional to the magnitude of the projection of the superlattice eigenstates on the primitive cell eigenstate. Up/down (red/blue) spin bands are split into separate frames to allow for clear visualization of otherwise overlapping states. Shaded pink/blue lines show the spin-orbit-split electronic bands of pristine MoS$_2$; these bands have been shifted in energy to align with the band edges of the unfolded superlattice.
}
\end{figure*}

We now investigate the influence of structural relaxation and strain-induced internal electric fields on the electronic bandstructure of monolayer MoS$_2$  and the emergence of superlattice minibands. As the external potential and structural relaxation can be viewed as perturbations to the pristine MoS$_2$  monolayer, unfolding the superlattice bandstructure  \cite{ku_unfolding_2010,popescu_extracting_2012,dirnberger_electronic_2021} from its reduced Brillouin zone to the full (extended) Brillouin zone of monolayer MoS$_2$  (Appendix B; Fig.\,6) allows for a more transparent analysis. Figures \ref{fig3}(A, B) display the unfolded bandstructures for the unrelaxed and relaxed armchair superlattices, from which we note several important distinctions. In the unrelaxed case (Fig. 3A), the change in band gap is small ($<1\%$; ~15 meV) for this magnitude of external potential ($V_0$=0.1V), and the gap remains direct at the K valleys. In the extended zone picture, we clearly observe the formation of superlattice minibands that are gapped due to the quantum confinement effect of the external potential. When the lattice is allowed to relax, these gapped minibands are further dispersed and the decrease in band gap is more substantial ($\sim 30\%$; 0.54 eV for $V_0$=0.1 V), the gap now being indirect between the $\Gamma$ valley at the valence band edge and the (degenerate) K valleys at the conduction band edge; the direct gap at the K valleys is $\sim 40$ meV larger. Similar results are seen in the zigzag case (Fig.\,S6). For a larger external potential ($V_0$=0.3 V; Fig.\,S7), the miniband dispersion becomes even more distinct with substantial miniband gaps (tens of meV). To demonstrate conclusively the structural basis of our findings, we recalculated the bandstructures of fully relaxed superlattices by artificially fixing atoms at their (field-dependent) positions and turning off the external potential (Figs.\,S8, S9) and found relatively small changes ($<10\%$) in the band gap with no qualitative changes in the miniband dispersion. We therefore conclude that the miniband dispersion occurs primarily due to Stark shifts driven by the internal polarization field that arises from piezoelectric coupling to the induced strain in the superlattice. Such effects are well known in quantum wells heterostructures composed of piezoelectric materials (e.g., zincblende-structure III-V superlattices \cite{smith_theory_1990}) in which strains at interfaces produce built-in electric fields that lead to Stark shifts and consequent dispersion of minibands that are accompanied by a decrease in the band gap. The important distinction is that, unlike quantum-well heterostructures involving dissimilar materials, we have here an electrostatic superlattice in a pristine monolayer of a single-phase 2D material that displays analogous physics. 

\begin{figure*}[htbp!]
\centering
\includegraphics[width=\linewidth]{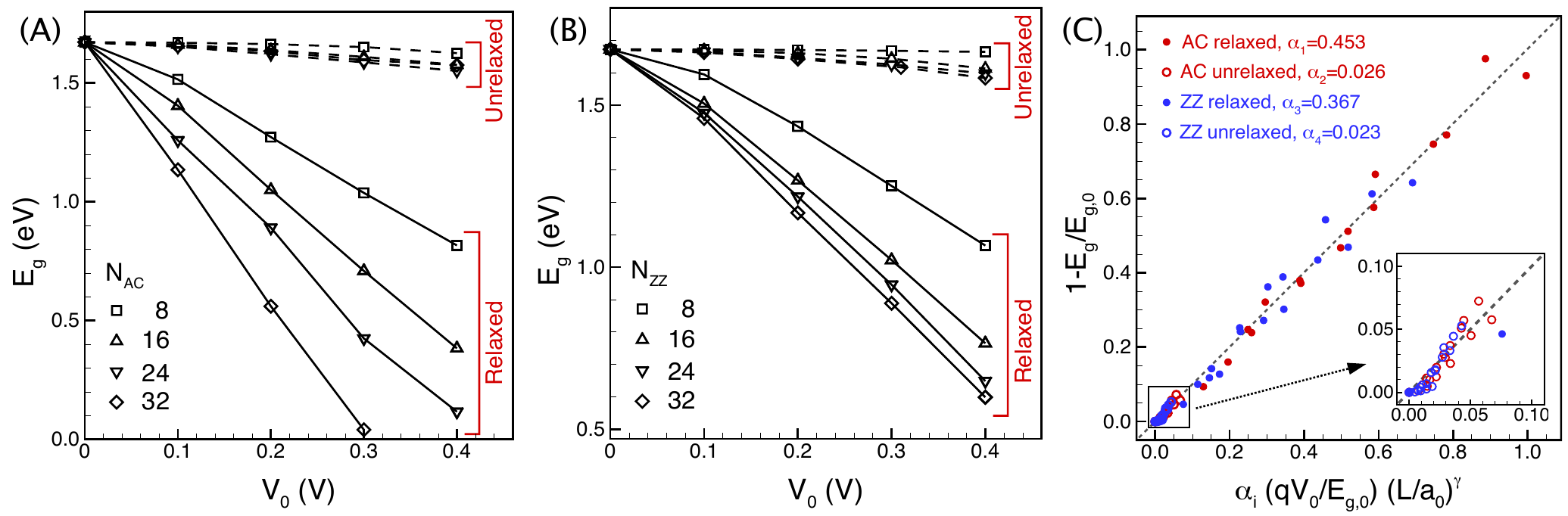}
\caption{\label{fig4} Bandgap ($E_g$) of monolayer MoS$_2$  subjected to a superlattice potential $V(\bm{r})=V_0\cos(\bm{q}\cdot\bm{r})$ along the armchair direction (A) and zigzag direction (B) with or without atomic relaxations. $N_{AC}$ and $N_{ZZ}$ are the number of repeat armchair or zigzag units in one period of the superlattice. The lines are a guide to the eye. (C) Parity plot of relative change in bandgap from DFT calculations versus power-law model with exponent  $\gamma=0.595$ ($R^2=0.99$).
}
\end{figure*}

To complete our analysis of bandgap tuning of monolayer MoS$_2$ electrostatic superlattices, we sampled the parameter space of such superlattices, including wavelengths and amplitudes of potential modulations as well as the effects of structural relaxation. The outcome of these studies is displayed in Figure \ref{fig4}, from which we see that: (i) for a given superlattice length, $L$, the band gap, $E_g$, decreases nearly linearly with the amplitude, $V_0$, of the applied potential;  \cite{ono_effect_2017} (ii) for a fixed value of $V_0$, a longer supercell experiences a greater reduction in band gap; and (iii) the reduction in band gap is substantially greater when structural relaxation is allowed. These observations hold irrespective of the direction – armchair or zigzag – of the potential modulation.  Based on these observations, we make a scaling ansatz 
\begin{equation}
E_g=E_{g,0} \left[ 1-\frac{\alpha q V_0}{E_{g,0}} \left( \frac{L}{a_0} \right)^\gamma\right], 
\end{equation}
where $E_{g,0}$ is the band gap of an unperturbed MoS$_2$  monolayer, $a_0$ is the lattice constant of the monolayer, $q$ is the elementary charge, and $\alpha$ and $\gamma$ are fitting parameters. Given the vast difference in behavior between unrelaxed and relaxed cases, as well as smaller but noticeable differences between the armchair and zigzag cases, we find it necessary to use different prefactors, $\alpha_i$ for each case; however, it is sufficient to use a single exponent $\gamma$ to fit the data accurately. For a specific direction (armchair/zigzag) of applied potential, a comparison of the prefactors, $\alpha_i$, for the relaxed and unrelaxed cases (Fig. \ref{fig4}C) indicates that the band gap is reduced by over an order of magnitude ($\sim$15 times) when superlattice relaxation is permitted. This unambiguously underscores the need to account for structural relaxation, which has been neglected in theoretical modeling of 2D electrostatic superlattices.
\footnote{In a practical device, one would expect the MoS2 monolayer to be partially constrained by the substrate, gate dielectric, contacts, etc., that preclude complete structural relaxation, the unrelaxed and (free-standing) fully relaxed cases serving as limiting cases.}
In contrast, the direction of the applied potential modulation is of less importance for bandgap tuning although it is significant for charge-carrier separation, as noted before. Finally, it is worth noting that the quantitative values of band gaps are underestimated due to the well-known limitations of semilocal exchange-correlation functionals, but the qualitative behavior presented here is expected to be robust (see Appendix C).

\subsection{Spin-valley coupling in minibands}

\begin{figure*}[htbp!]
\centering
\includegraphics[width=\linewidth]{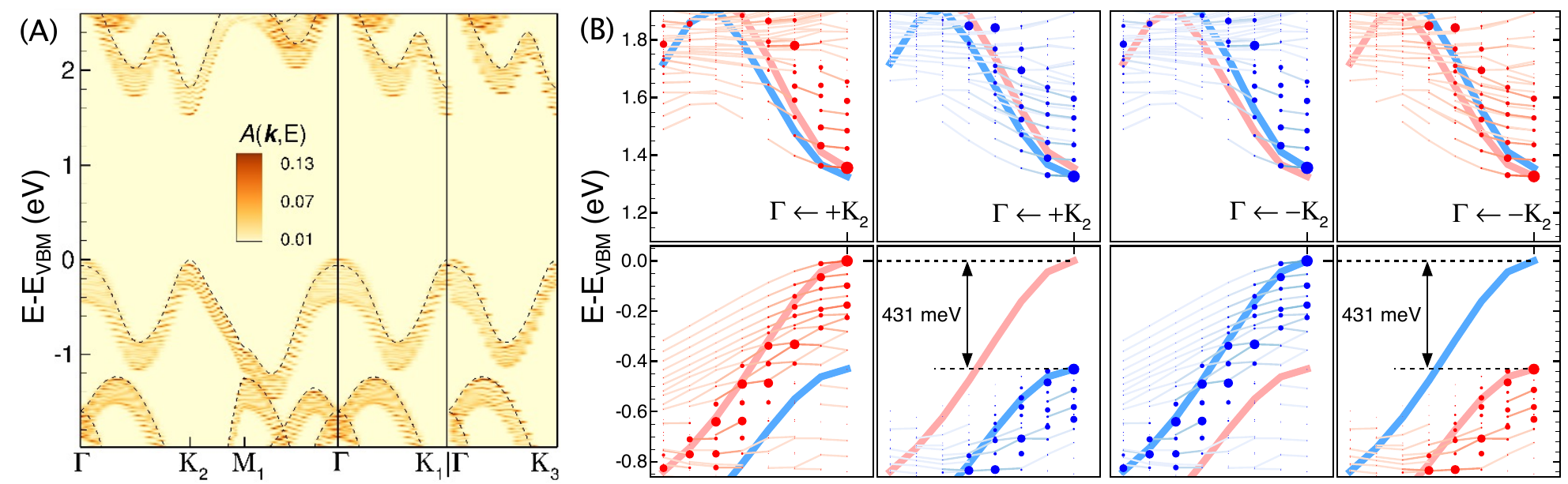}
\caption{\label{fig5} (A) Unfolding into the first Brillouin zone of the electronic bandstructure of a fully relaxed WS$_2$ superlattice ($N_{AC}$=32) subjected to an external potential $V(\bm{r})=V_0\cos(\bm{q}_{AC}\cdot\bm{r})$ with $V_0=0.1$V. Bands of the pristine (unperturbed) WS$_2$ monolayer are overlaid in thin dashed lines. The intensity of the color map indicates the overall contribution from supercell eigenstates to a primitive cell eigenstate at wavevector $\bm{k}$ and energy $E$ – the spectral function $A(\bm{k}, E)$.  (B) Details of the unfolded electronic bandstructure at the $\pm K_2$ valleys with the inclusion of spin-orbit splitting. Symbols are colored by the degree of spin polarization ($\eta_z=m_z/\left|m\right|$); the size of the symbol is proportional to the magnitude of the projection of the superlattice eigenstates on the primitive cell eigenstate. Up/down (red/blue) spin bands are split into separate frames to allow for clear visualization of otherwise overlapping states. Shaded pink/blue lines show the spin-orbit-split electronic bands of pristine WS$_2$; these bands have been shifted in energy to align with the band edges of the unfolded superlattice.
}
\end{figure*}

One of the most interesting properties of monolayer MoS$_2$  is the coupling between valley and spin degrees of freedom, due the combination of strong spin-orbit splitting at the valence band edge and lack of centrosymmetry. \cite{xiao_coupled_2012,mak_control_2012} Specifically, in the absence of magnetic fields, the $K$ and $-K$ valleys of monolayer MoS$_2$ are related only by time-reversal symmetry and associated with distinct and opposite magnetic moments. Thus, it is natural to inquire whether these desirable valley-selective properties are still preserved in electrostatic superlattices, in spite of large changes in electronic band gaps and, more importantly, significant Stark shifts at the band edges. Figures \ref{fig3}(C, D) display the spin-orbit-split band edges at the $\pm K_2$ valleys of the corresponding bandstructures displayed in Figures \ref{fig3}(A, B) respectively. In the unrelaxed case (Fig.\,ref{fig3}C), which we recall presents a very small change in band gap and no significant miniband dispersion, the superlattice states essentially follow the dispersion of the primitive cell eigenstates and preserve the valley polarization. The situation is more interesting for the relaxed case (Fig. \ref{fig3}(D)), which shows the formation of distinct spin-polarized minibands that preserve the strong spin-polarization ($m_z\to\pm1$) of the $\pm K_2$ valleys. At the valence band edge, in particular, minibands of exclusively one spin-polarization populate the spin-orbit gap ($\sim$150 meV \cite{zhu_giant_2011,ramasubramaniam_large_2012}) and, importantly, these minibands are gapped by several tens of meVs. At the conduction band edge, the spin-orbit splitting is very small \cite{kormanyos_monolayer_2013, marinov_resolving_2017} and, while the miniband gaps are appreciable, the up/down spin states are nearly degenerate. Increasing the potential amplitude ($V_0$=0.3V; Fig.\,S9) further increases the miniband splitting to $\sim$100 meV but there still remains a single spin-polarized miniband within the spin-orbit gap, suggesting a degree of tunability of these miniband energies. The zigzag superlattice, in contrast, does not show significant miniband dispersion at the valence band edge of the $K$ valleys (Fig.\,S6) and thus, there are no miniband states within the spin-orbit gap.

While valley-polarization and the consequent valley-selective properties of monolayer MoS$_2$ (and several other 2D materials) are well known, the possibility of engineering valley-polarized minibands within a single-phase monolayer is of particular significance. Not only are minibands of importance for optical emission/detection in the mid-infrared and THz spectral ranges, but the fact that these minibands are valley-polarized implies that the such emission/detection could also be spin-selective, enabling future spintronic and quantum electronics applications. Furthermore, while MoS$_2$ already presents an appreciable spin-orbit gap ($\sim$150 meV), opening up this gap further could provide an even larger energy window within which miniband splitting can be tuned. To this end, we display in Figure \ref{fig5}, an example of the bandstructure of a relaxed superlattice of monolayer WS$_2$ -- also a piezoelectric material \cite{duerloo_intrinsic_2012} -- subjected to a cosine potential of amplitude $V_0$=0.1V along the armchair direction (in exact correspondence with the MoS$_2$ case). Spin-orbit splitting of the valence band edge at the K valleys in monolayer WS$_2$ is nearly three times larger than in MoS$_2$  \cite{zhu_giant_2011,ramasubramaniam_large_2012}, and we now observe several spin-polarized minibands; increasing the potential to $V_0$=0.3V further enhances the miniband splitting (Fig.\,S10). Thus, we suggest that electrostatic superlattices in piezoelectric 2D materials \cite{cui_two-dimensional_2018} with strong spin-orbit coupling can add a new dimension of miniband engineering to ongoing efforts in quantum electronics.

\section{Conclusions}
In conclusion, we explored 1D superlattices induced by a confining electrostatic potential in monolayer MoS$_2$, a prototypical two-dimensional (2D) semiconductor, and demonstrated via first-principles calculations that the electronic response of the material to the applied potential is dominated by field-induced structural distortions rather than pure shifts in electronic levels alone. These structural distortions are tunable with applied potential and reduce the intrinsic band gap of monolayer MoS$_2$ substantially while also polarizing the monolayer through piezoelectric coupling, resulting in spatial separation of charge carriers as well as large Stark shifts that produce well-separated (several 10s of meV) dispersive minibands that inherit the valley-selective magnetic properties. Consequently, we suggest that coupling between electric fields and internal strains in monolayer MoS$_2$ and analogous TMDCs could be harnessed to design continuously tunable spin-selective emitters/detectors in the mid-infrared and THz spectral ranges, enabling further application of these materials in spintronics and quantum electronics.

\section*{Acknowledgements} 
We gratefully acknowledge the National Science Foundation (NSF-BSF 2150562) and the US-Israel Binational Science Foundation (2021721) for support. This work used Bridges2 at the Pittsburgh Supercomputing Center through allocation TG-DMR190070 from the Advanced Cyberinfrastructure Coordination Ecosystem: Services \& Support (ACCESS) program, which is supported by National Science Foundation grants \#2138259, \#2138286, \#2138307, \#2137603, and \#2138296. Research computing support from the Office of Information Technology at the University of Massachusetts Amherst is also gratefully acknowledged.\vspace{0.2cm}

\appendix
\section{Density Functional Theory Modeling}

\begin{figure}[h!]
\centering
\includegraphics[width=\linewidth]{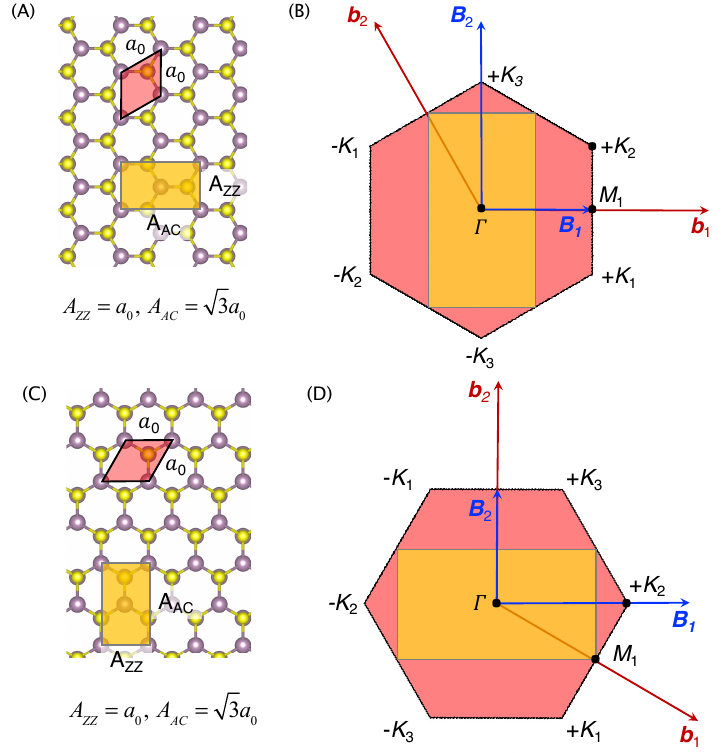}
\caption{\label{fig6} (A, C) Primitive (red rhombus) and orthorhombic (yellow rectangle) unit cells of MoS2; black and yellow spheres indicate Mo and S atoms. Orthorhombic unit cells are repeated $N_{AC}$ or $N_{ZZ}$ times along the horizontal direction in panels (A) or (C), respectively, to produce simulation supercells. The simulation supercells are fully periodic in-plane. The lattice parameter for the (relaxed) primitive cell is $a_0=3.183$\AA. (B, D) First Brillouin zones of the primitive and orthorhombic unit cells, colored in correspondence to their real-space counterparts, and their reciprocal lattice vectors indicated in lower- and upper-case, respectively. Select high-symmetry k-points of the first Brillouin zone of the primitive cell are also noted; the K points are denoted as individual time-reversal-symmetric ($\pm$) pairs, as their nominal threefold symmetry could be broken by the applied external potential. 
}
\end{figure}

Density functional theory (DFT) modeling was performed using the Vienna Ab Initio Simulation Package (VASP; version 6.2.1). \cite{kresse_efficient_1996,kresse_efficiency_1996}  The projector-augmented wave method \cite{blochl_projector_1994,kresse_ultrasoft_1999} was used to represent core and valence electrons; the valence electronic configurations are $5s^14d^5$ for Mo, $6s^15d^5$ for W, and $3s^2 3p^4$ for S. Electron exchange and correlation were modeled using the Perdew-Burke-Ernzerhof form of the generalized-gradient approximation. \cite{perdew_generalized_1996} The kinetic energy cutoff for planewaves was set to 400 eV. Gaussian smearing of 0.01 eV was used for Brillouin zone integrations in conjunction with a dense grid of 3000 points to sample the density of states. Orthorhombic unit cells of 2H MoS$_2$ and WS$_2$ monolayers were fully relaxed (both atomic positions and lattice vectors) using the conjugate-gradient method with a force tolerance of 0.01 eV/\AA~ and energy tolerance of $10^{-4}$ eV. Supercells were thereafter constructed by replicating these relaxed unit cells along the armchair or zigzag directions (Fig.\ref{fig6}) and subsequently only atomic positions of these supercells were relaxed under applied external fields. $\Gamma$-centered $k$-point meshes for relaxation calculations were constructed using 5 and 8 grid points, respectively, along the shorter and longer reciprocal lattice vectors (Fig.\,\ref{fig6}) of the orthorhombic unit cell and these grid points were proportionately decreased for the supercells. Subsequent to structural relaxation, the $k$-point mesh density was doubled to obtain more accurate electronic structures for the various supercells. To render the calculations tractable, spin-orbit coupling was neglected in the relaxation and single-point calculations; this has no bearing on the qualitative trends and overall conclusions drawn in this work. A few select examples were studied with the inclusion of spin-orbit coupling to demonstrate the impact of the superlattice potential on spin–valley coupling, as discussed in the main text.

The VASP source code was modified to apply an external periodic potential, $V(\bm r)=V_0\cos(\bm q \cdot \bm r)$ by customizing the existing subroutine EXTERNAL\_POT in the pot.F file.

\section{Unfolding Calculations}
Brillouin zone unfolding was performed using the bands4vasp postprocessing package (\url{https://github.com/QuantumMaterialsModelling/bands4vasp}) with the unfolding patch (\url{https://github.com/QuantumMaterialsModelling/UnfoldingPatch4vasp}) applied to the VASP source code. \cite{dirnberger_electronic_2021} To unfold the supercell Brillouin zone into that of the primitive cell, it is necessary that the supercell lattice vectors ($\bf A$) and the primitive cell lattice vectors ($\bf a$) satisfy the relation  $\bm{A=Ma}$, where all elements of the transformation matrix, $\bm M$, are integers. \cite{dirnberger_electronic_2021} As an example, consider the primitive cell and orthorhombic unit cell in Figure \ref{fig6}(A): the corresponding cell vectors are $\bm{A}_1=[\sqrt{3}a_0,0], \bm{A}_2=[0,a_0], \bm{a}_1=[\sqrt{3}a_0/2,a_0/2], \bm{a}_2=[0,a_0]$. It is straightforward to show that the integer transformation matrix is
\begin{equation}
\bm{M}=\left[ \begin{array} {cc} 2 & -1 \\ 0 & 1 \end{array} \right]
\end{equation}
and, furthermore, if the orthogonal unit cell is replicated $N_{AC}$ times along the armchair direction to produce the supercell, the transformation matrix becomes 
\begin{equation}
\bm{M}=\left[ \begin{array} {cc} 2N_{AC} & -N_{AC} \\ 0 & 1 \end{array} \right].
\end{equation}
Similarly, for the choice of primitive and orthogonal cells in Fig.\,\ref{fig6}(B), the transformation matrix is 
\begin{equation}
\bm{M}=\left[ \begin{array} {cc} 1 & 0 \\ -1 & 2 \end{array} \right]
\end{equation}
and, if the orthogonal unit cell is replicated $N_{ZZ}$ times along the zigzag direction, the transformation matrix becomes 
\begin{equation}
\bm{M}=\left[ \begin{array} {cc} N_{ZZ} & 0 \\ -1 & 2 \end{array} \right].
\end{equation}
We use these two sets of supercells and transformation matrices in our calculations. Electronic bandgaps are obtained by unfolding the superlattice bandstructure along the $\Gamma-K_2$ line (Fig.\,\ref{fig6}) and calculating the smallest gap, which is typically either direct at $K_2$ or indirect between $\Gamma$ at the valence band edge and $K_2$ at the conduction band edge.

\section{Comparison of semilocal and hybrid functionals}
\begin{figure*}[h!]
\centering
\includegraphics[width=0.8\linewidth]{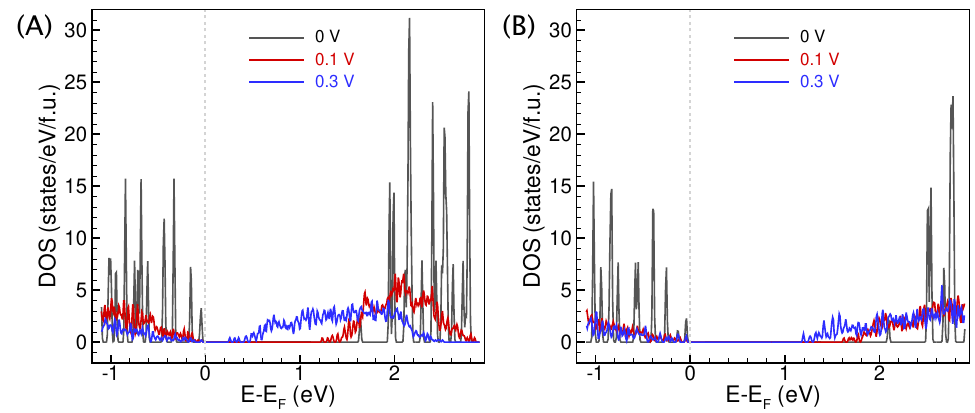}
\caption{\label{fig7} Density of states (DOS) per formula unit (f.u.) of MoS$_2$ using the (A) PBE and (B) HSE functionals. The black line indicates the DOS of the unperturbed MoS$_2$ monolayer (primitive cell); red and blue lines correspond to the relaxed armchair superlattices ($N_{AC}=32$) with potential amplitude $V_0$ as indicated in the plot legend.
}
\end{figure*}

Semilocal exchange-correlation functionals such as PBE are well known to underestimate band gaps in materials. While calculation of quantitively accurate band gaps of electrostatic superlattices might potentially be accomplished using more accurate functionals \cite{ramasubramaniam_transferable_2019}---albeit, at higher computational cost---the main purpose of this paper is to demonstrate qualitative features of piezoelectric coupling and miniband formation in electrostatic superlattices. Nevertheless, it is worth confirming that the results are not artifacts of the choice of the exchange-correlation functional and, to this end, we performed a few test calculations on an armchair superlattice ($N_{AC}=32$), employing the HSE hybrid functional \cite{krukau_influence_2006} with PBE-relaxed structures as inputs. As hybrid-DFT calculations of these superlattices (192 atoms for $N_{AC}=32$) are computationally demanding, we restricted calculation of the Fock operator (exact exchange contribution) to just the $\Gamma$-point: the calculations are thus, not fully converged at the HSE level but the outcome is nevertheless sufficient to corroborate the PBE results. No down-sampling was needed for the primitive cell. Figure \ref{fig7} displays the density of states (DOS) calculated with PBE and HSE functionals for a couple of different potential amplitudes ($V_0$). As evident, the band gap clearly decreases with increasing magnitude of the applied potential. Furthermore, the sharp resonances in the DOS of the unperturbed monolayer are smeared out in the (relaxed) superlattices and we note the emergence of smaller peaks near the band edges, consistent with the appearance of superlattice minibands.

\nolinenumbers

\begin{thebibliography}{69}%
\makeatletter
\providecommand \@ifxundefined [1]{%
 \@ifx{#1\undefined}
}%
\providecommand \@ifnum [1]{%
 \ifnum #1\expandafter \@firstoftwo
 \else \expandafter \@secondoftwo
 \fi
}%
\providecommand \@ifx [1]{%
 \ifx #1\expandafter \@firstoftwo
 \else \expandafter \@secondoftwo
 \fi
}%
\providecommand \natexlab [1]{#1}%
\providecommand \enquote  [1]{``#1''}%
\providecommand \bibnamefont  [1]{#1}%
\providecommand \bibfnamefont [1]{#1}%
\providecommand \citenamefont [1]{#1}%
\providecommand \href@noop [0]{\@secondoftwo}%
\providecommand \href [0]{\begingroup \@sanitize@url \@href}%
\providecommand \@href[1]{\@@startlink{#1}\@@href}%
\providecommand \@@href[1]{\endgroup#1\@@endlink}%
\providecommand \@sanitize@url [0]{\catcode `\\12\catcode `\$12\catcode
  `\&12\catcode `\#12\catcode `\^12\catcode `\_12\catcode `\%12\relax}%
\providecommand \@@startlink[1]{}%
\providecommand \@@endlink[0]{}%
\providecommand \url  [0]{\begingroup\@sanitize@url \@url }%
\providecommand \@url [1]{\endgroup\@href {#1}{\urlprefix }}%
\providecommand \urlprefix  [0]{URL }%
\providecommand \Eprint [0]{\href }%
\providecommand \doibase [0]{https://doi.org/}%
\providecommand \selectlanguage [0]{\@gobble}%
\providecommand \bibinfo  [0]{\@secondoftwo}%
\providecommand \bibfield  [0]{\@secondoftwo}%
\providecommand \translation [1]{[#1]}%
\providecommand \BibitemOpen [0]{}%
\providecommand \bibitemStop [0]{}%
\providecommand \bibitemNoStop [0]{.\EOS\space}%
\providecommand \EOS [0]{\spacefactor3000\relax}%
\providecommand \BibitemShut  [1]{\csname bibitem#1\endcsname}%
\let\auto@bib@innerbib\@empty
\bibitem [{\citenamefont {Esaki}\ and\ \citenamefont
  {Tsu}(1970)}]{esaki_superlattice_1970}%
  \BibitemOpen
  \bibfield  {author} {\bibinfo {author} {\bibfnamefont {L.}~\bibnamefont
  {Esaki}}\ and\ \bibinfo {author} {\bibfnamefont {R.}~\bibnamefont {Tsu}},\
  }\bibfield  {title} {\bibinfo {title} {Superlattice and {Negative}
  {Differential} {Conductivity} in {Semiconductors}},\ }\href
  {https://doi.org/10.1147/rd.141.0061} {\bibfield  {journal} {\bibinfo
  {journal} {IBM Journal of Research and Development}\ }\textbf {\bibinfo
  {volume} {14}},\ \bibinfo {pages} {61} (\bibinfo {year} {1970})}\BibitemShut
  {NoStop}%
\bibitem [{\citenamefont {Smith}\ and\ \citenamefont
  {Mailhiot}(1990)}]{smith_theory_1990}%
  \BibitemOpen
  \bibfield  {author} {\bibinfo {author} {\bibfnamefont {D.~L.}\ \bibnamefont
  {Smith}}\ and\ \bibinfo {author} {\bibfnamefont {C.}~\bibnamefont
  {Mailhiot}},\ }\bibfield  {title} {\bibinfo {title} {Theory of semiconductor
  superlattice electronic structure},\ }\href
  {https://doi.org/10.1103/RevModPhys.62.173} {\bibfield  {journal} {\bibinfo
  {journal} {Rev. Mod. Phys.}\ }\textbf {\bibinfo {volume} {62}},\ \bibinfo
  {pages} {173} (\bibinfo {year} {1990})}\BibitemShut {NoStop}%
\bibitem [{\citenamefont {Forsythe}\ \emph {et~al.}(2018)\citenamefont
  {Forsythe}, \citenamefont {Zhou}, \citenamefont {Watanabe}, \citenamefont
  {Taniguchi}, \citenamefont {Pasupathy}, \citenamefont {Moon}, \citenamefont
  {Koshino}, \citenamefont {Kim},\ and\ \citenamefont
  {Dean}}]{forsythe_band_2018}%
  \BibitemOpen
  \bibfield  {author} {\bibinfo {author} {\bibfnamefont {C.}~\bibnamefont
  {Forsythe}}, \bibinfo {author} {\bibfnamefont {X.}~\bibnamefont {Zhou}},
  \bibinfo {author} {\bibfnamefont {K.}~\bibnamefont {Watanabe}}, \bibinfo
  {author} {\bibfnamefont {T.}~\bibnamefont {Taniguchi}}, \bibinfo {author}
  {\bibfnamefont {A.}~\bibnamefont {Pasupathy}}, \bibinfo {author}
  {\bibfnamefont {P.}~\bibnamefont {Moon}}, \bibinfo {author} {\bibfnamefont
  {M.}~\bibnamefont {Koshino}}, \bibinfo {author} {\bibfnamefont
  {P.}~\bibnamefont {Kim}},\ and\ \bibinfo {author} {\bibfnamefont {C.~R.}\
  \bibnamefont {Dean}},\ }\bibfield  {title} {\bibinfo {title} {Band structure
  engineering of {2D} materials using patterned dielectric superlattices},\
  }\href@noop {} {\bibfield  {journal} {\bibinfo  {journal} {Nat. Nanotech.}\
  }\textbf {\bibinfo {volume} {13}},\ \bibinfo {pages} {566} (\bibinfo {year}
  {2018})}\BibitemShut {NoStop}%
\bibitem [{\citenamefont {Li}\ \emph {et~al.}(2021)\citenamefont {Li},
  \citenamefont {Dietrich}, \citenamefont {Forsythe}, \citenamefont
  {Taniguchi}, \citenamefont {Watanabe}, \citenamefont {Moon},\ and\
  \citenamefont {Dean}}]{li_anisotropic_2021}%
  \BibitemOpen
  \bibfield  {author} {\bibinfo {author} {\bibfnamefont {Y.}~\bibnamefont
  {Li}}, \bibinfo {author} {\bibfnamefont {S.}~\bibnamefont {Dietrich}},
  \bibinfo {author} {\bibfnamefont {C.}~\bibnamefont {Forsythe}}, \bibinfo
  {author} {\bibfnamefont {T.}~\bibnamefont {Taniguchi}}, \bibinfo {author}
  {\bibfnamefont {K.}~\bibnamefont {Watanabe}}, \bibinfo {author}
  {\bibfnamefont {P.}~\bibnamefont {Moon}},\ and\ \bibinfo {author}
  {\bibfnamefont {C.~R.}\ \bibnamefont {Dean}},\ }\bibfield  {title} {\bibinfo
  {title} {Anisotropic band flattening in graphene with one-dimensional
  superlattices},\ }\href {https://doi.org/10.1038/s41565-021-00849-9}
  {\bibfield  {journal} {\bibinfo  {journal} {Nat. Nanotechnol.}\ ,\ \bibinfo
  {pages} {1}} (\bibinfo {year} {2021})}\BibitemShut {NoStop}%
\bibitem [{\citenamefont {Huber}\ \emph {et~al.}(2020)\citenamefont {Huber},
  \citenamefont {Liu}, \citenamefont {Chen}, \citenamefont {Drienovsky},
  \citenamefont {Sandner}, \citenamefont {Watanabe}, \citenamefont {Taniguchi},
  \citenamefont {Richter}, \citenamefont {Weiss},\ and\ \citenamefont
  {Eroms}}]{huber_gate-tunable_2020}%
  \BibitemOpen
  \bibfield  {author} {\bibinfo {author} {\bibfnamefont {R.}~\bibnamefont
  {Huber}}, \bibinfo {author} {\bibfnamefont {M.-H.}\ \bibnamefont {Liu}},
  \bibinfo {author} {\bibfnamefont {S.-C.}\ \bibnamefont {Chen}}, \bibinfo
  {author} {\bibfnamefont {M.}~\bibnamefont {Drienovsky}}, \bibinfo {author}
  {\bibfnamefont {A.}~\bibnamefont {Sandner}}, \bibinfo {author} {\bibfnamefont
  {K.}~\bibnamefont {Watanabe}}, \bibinfo {author} {\bibfnamefont
  {T.}~\bibnamefont {Taniguchi}}, \bibinfo {author} {\bibfnamefont
  {K.}~\bibnamefont {Richter}}, \bibinfo {author} {\bibfnamefont
  {D.}~\bibnamefont {Weiss}},\ and\ \bibinfo {author} {\bibfnamefont
  {J.}~\bibnamefont {Eroms}},\ }\bibfield  {title} {\bibinfo {title}
  {Gate-{Tunable} {Two}-{Dimensional} {Superlattices} in {Graphene}},\
  }\href@noop {} {\bibfield  {journal} {\bibinfo  {journal} {Nano Lett.}\
  }\textbf {\bibinfo {volume} {20}},\ \bibinfo {pages} {8046} (\bibinfo {year}
  {2020})}\BibitemShut {NoStop}%
\bibitem [{\citenamefont {Barcons~Ruiz}\ \emph {et~al.}(2022)\citenamefont
  {Barcons~Ruiz}, \citenamefont {Herzig~Sheinfux}, \citenamefont {Hoffmann},
  \citenamefont {Torre}, \citenamefont {Agarwal}, \citenamefont {Kumar},
  \citenamefont {Vistoli}, \citenamefont {Taniguchi}, \citenamefont {Watanabe},
  \citenamefont {Bachtold},\ and\ \citenamefont
  {Koppens}}]{barcons_ruiz_engineering_2022}%
  \BibitemOpen
  \bibfield  {author} {\bibinfo {author} {\bibfnamefont {D.}~\bibnamefont
  {Barcons~Ruiz}}, \bibinfo {author} {\bibfnamefont {H.}~\bibnamefont
  {Herzig~Sheinfux}}, \bibinfo {author} {\bibfnamefont {R.}~\bibnamefont
  {Hoffmann}}, \bibinfo {author} {\bibfnamefont {I.}~\bibnamefont {Torre}},
  \bibinfo {author} {\bibfnamefont {H.}~\bibnamefont {Agarwal}}, \bibinfo
  {author} {\bibfnamefont {R.~K.}\ \bibnamefont {Kumar}}, \bibinfo {author}
  {\bibfnamefont {L.}~\bibnamefont {Vistoli}}, \bibinfo {author} {\bibfnamefont
  {T.}~\bibnamefont {Taniguchi}}, \bibinfo {author} {\bibfnamefont
  {K.}~\bibnamefont {Watanabe}}, \bibinfo {author} {\bibfnamefont
  {A.}~\bibnamefont {Bachtold}},\ and\ \bibinfo {author} {\bibfnamefont
  {F.~H.~L.}\ \bibnamefont {Koppens}},\ }\bibfield  {title} {\bibinfo {title}
  {Engineering high quality graphene superlattices via ion milled ultra-thin
  etching masks},\ }\href@noop {} {\bibfield  {journal} {\bibinfo  {journal}
  {Nat. Commun.}\ }\textbf {\bibinfo {volume} {13}},\ \bibinfo {pages} {6926}
  (\bibinfo {year} {2022})}\BibitemShut {NoStop}%
\bibitem [{\citenamefont {Dragoman}\ \emph {et~al.}(2021)\citenamefont
  {Dragoman}, \citenamefont {Dinescu}, \citenamefont {Dragoman},\ and\
  \citenamefont {Comanescu}}]{dragoman_bloch_2021}%
  \BibitemOpen
  \bibfield  {author} {\bibinfo {author} {\bibfnamefont {M.}~\bibnamefont
  {Dragoman}}, \bibinfo {author} {\bibfnamefont {A.}~\bibnamefont {Dinescu}},
  \bibinfo {author} {\bibfnamefont {D.}~\bibnamefont {Dragoman}},\ and\
  \bibinfo {author} {\bibfnamefont {F.}~\bibnamefont {Comanescu}},\ }\bibfield
  {title} {\bibinfo {title} {Bloch oscillations at room temperature in
  graphene/h-bn electrostatic superlattices},\ }\href@noop {} {\bibfield
  {journal} {\bibinfo  {journal} {Nanotechnology}\ }\textbf {\bibinfo {volume}
  {32}},\ \bibinfo {pages} {345203} (\bibinfo {year} {2021})}\BibitemShut
  {NoStop}%
\bibitem [{\citenamefont {Kyoung Ryu}\ \emph {et~al.}(2019)\citenamefont
  {Kyoung Ryu}, \citenamefont {Frisenda},\ and\ \citenamefont
  {Castellanos-Gomez}}]{kyoungryu_superlattices_2019}%
  \BibitemOpen
  \bibfield  {author} {\bibinfo {author} {\bibfnamefont {Y.}~\bibnamefont
  {Kyoung Ryu}}, \bibinfo {author} {\bibfnamefont {R.}~\bibnamefont
  {Frisenda}},\ and\ \bibinfo {author} {\bibfnamefont {A.}~\bibnamefont
  {Castellanos-Gomez}},\ }\bibfield  {title} {\bibinfo {title} {Superlattices
  based on van der {Waals} {2D} materials},\ }\href
  {https://doi.org/10.1039/C9CC04919C} {\bibfield  {journal} {\bibinfo
  {journal} {Chemical Communications}\ }\textbf {\bibinfo {volume} {55}},\
  \bibinfo {pages} {11498} (\bibinfo {year} {2019})}\BibitemShut {NoStop}%
\bibitem [{\citenamefont {Schmeller}\ \emph {et~al.}(1994)\citenamefont
  {Schmeller}, \citenamefont {Hansen}, \citenamefont {Kotthaus}, \citenamefont
  {Tränkle},\ and\ \citenamefont {Weimann}}]{schmeller_franzkeldysh_1994}%
  \BibitemOpen
  \bibfield  {author} {\bibinfo {author} {\bibfnamefont {A.}~\bibnamefont
  {Schmeller}}, \bibinfo {author} {\bibfnamefont {W.}~\bibnamefont {Hansen}},
  \bibinfo {author} {\bibfnamefont {J.~P.}\ \bibnamefont {Kotthaus}}, \bibinfo
  {author} {\bibfnamefont {G.}~\bibnamefont {Tränkle}},\ and\ \bibinfo
  {author} {\bibfnamefont {G.}~\bibnamefont {Weimann}},\ }\bibfield  {title}
  {\bibinfo {title} {Franz–{Keldysh} effect in a two‐dimensional system},\
  }\href {https://doi.org/10.1063/1.111166} {\bibfield  {journal} {\bibinfo
  {journal} {Appl. Phys. Lett.}\ }\textbf {\bibinfo {volume} {64}},\ \bibinfo
  {pages} {330} (\bibinfo {year} {1994})}\BibitemShut {NoStop}%
\bibitem [{\citenamefont {Schlösser}\ \emph {et~al.}(1996)\citenamefont
  {Schlösser}, \citenamefont {Ensslin}, \citenamefont {Kotthaus},\ and\
  \citenamefont {Holland}}]{schlosser_landau_1996}%
  \BibitemOpen
  \bibfield  {author} {\bibinfo {author} {\bibfnamefont {T.}~\bibnamefont
  {Schlösser}}, \bibinfo {author} {\bibfnamefont {K.}~\bibnamefont {Ensslin}},
  \bibinfo {author} {\bibfnamefont {J.~P.}\ \bibnamefont {Kotthaus}},\ and\
  \bibinfo {author} {\bibfnamefont {M.}~\bibnamefont {Holland}},\ }\bibfield
  {title} {\bibinfo {title} {Landau subbands generated by a lateral
  electrostatic superlattice - chasing the {Hofstadter} butterfly},\ }\href
  {https://doi.org/10.1088/0268-1242/11/11S/022} {\bibfield  {journal}
  {\bibinfo  {journal} {Semicond. Sci. Technol.}\ }\textbf {\bibinfo {volume}
  {11}},\ \bibinfo {pages} {1582} (\bibinfo {year} {1996})}\BibitemShut
  {NoStop}%
\bibitem [{\citenamefont {Zimmermann}\ \emph {et~al.}(1997)\citenamefont
  {Zimmermann}, \citenamefont {Govorov}, \citenamefont {Hansen}, \citenamefont
  {Kotthaus}, \citenamefont {Bichler},\ and\ \citenamefont
  {Wegscheider}}]{zimmermann_lateral_1997}%
  \BibitemOpen
  \bibfield  {author} {\bibinfo {author} {\bibfnamefont {S.}~\bibnamefont
  {Zimmermann}}, \bibinfo {author} {\bibfnamefont {A.~O.}\ \bibnamefont
  {Govorov}}, \bibinfo {author} {\bibfnamefont {W.}~\bibnamefont {Hansen}},
  \bibinfo {author} {\bibfnamefont {J.~P.}\ \bibnamefont {Kotthaus}}, \bibinfo
  {author} {\bibfnamefont {M.}~\bibnamefont {Bichler}},\ and\ \bibinfo {author}
  {\bibfnamefont {W.}~\bibnamefont {Wegscheider}},\ }\bibfield  {title}
  {\bibinfo {title} {Lateral superlattices as voltage-controlled traps for
  excitons},\ }\href {https://doi.org/10.1103/PhysRevB.56.13414} {\bibfield
  {journal} {\bibinfo  {journal} {Phys. Rev. B}\ }\textbf {\bibinfo {volume}
  {56}},\ \bibinfo {pages} {13414} (\bibinfo {year} {1997})}\BibitemShut
  {NoStop}%
\bibitem [{\citenamefont {Yankowitz}\ \emph {et~al.}(2019)\citenamefont
  {Yankowitz}, \citenamefont {Ma}, \citenamefont {Jarillo-Herrero},\ and\
  \citenamefont {LeRoy}}]{yankowitz_van_2019}%
  \BibitemOpen
  \bibfield  {author} {\bibinfo {author} {\bibfnamefont {M.}~\bibnamefont
  {Yankowitz}}, \bibinfo {author} {\bibfnamefont {Q.}~\bibnamefont {Ma}},
  \bibinfo {author} {\bibfnamefont {P.}~\bibnamefont {Jarillo-Herrero}},\ and\
  \bibinfo {author} {\bibfnamefont {B.~J.}\ \bibnamefont {LeRoy}},\ }\bibfield
  {title} {\bibinfo {title} {van der {Waals} heterostructures combining
  graphene and hexagonal boron nitride},\ }\href@noop {} {\bibfield  {journal}
  {\bibinfo  {journal} {Nat. Rev. Phys.}\ }\textbf {\bibinfo {volume} {1}},\
  \bibinfo {pages} {112} (\bibinfo {year} {2019})}\BibitemShut {NoStop}%
\bibitem [{\citenamefont {Choi}\ \emph {et~al.}(2014)\citenamefont {Choi},
  \citenamefont {Park},\ and\ \citenamefont {Louie}}]{choi_electron_2014}%
  \BibitemOpen
  \bibfield  {author} {\bibinfo {author} {\bibfnamefont {S.}~\bibnamefont
  {Choi}}, \bibinfo {author} {\bibfnamefont {C.-H.}\ \bibnamefont {Park}},\
  and\ \bibinfo {author} {\bibfnamefont {S.~G.}\ \bibnamefont {Louie}},\
  }\bibfield  {title} {\bibinfo {title} {Electron {Supercollimation} in
  {Graphene} and {Dirac} {Fermion} {Materials} {Using} {One}-{Dimensional}
  {Disorder} {Potentials}},\ }\href
  {https://doi.org/10.1103/PhysRevLett.113.026802} {\bibfield  {journal}
  {\bibinfo  {journal} {Phys. Rev. Lett.}\ }\textbf {\bibinfo {volume} {113}},\
  \bibinfo {pages} {026802} (\bibinfo {year} {2014})}\BibitemShut {NoStop}%
\bibitem [{\citenamefont {Balents}\ \emph {et~al.}(2020)\citenamefont
  {Balents}, \citenamefont {Dean}, \citenamefont {Efetov},\ and\ \citenamefont
  {Young}}]{balents_superconductivity_2020}%
  \BibitemOpen
  \bibfield  {author} {\bibinfo {author} {\bibfnamefont {L.}~\bibnamefont
  {Balents}}, \bibinfo {author} {\bibfnamefont {C.~R.}\ \bibnamefont {Dean}},
  \bibinfo {author} {\bibfnamefont {D.~K.}\ \bibnamefont {Efetov}},\ and\
  \bibinfo {author} {\bibfnamefont {A.~F.}\ \bibnamefont {Young}},\ }\bibfield
  {title} {\bibinfo {title} {Superconductivity and strong correlations in
  moiré flat bands},\ }\href@noop {} {\bibfield  {journal} {\bibinfo
  {journal} {Nat. Phys.}\ }\textbf {\bibinfo {volume} {16}},\ \bibinfo {pages}
  {725} (\bibinfo {year} {2020})}\BibitemShut {NoStop}%
\bibitem [{\citenamefont {Behura}\ \emph {et~al.}(2021)\citenamefont {Behura},
  \citenamefont {Miranda}, \citenamefont {Nayak}, \citenamefont {Johnson},
  \citenamefont {Das},\ and\ \citenamefont {Pradhan}}]{behura_moire_2021}%
  \BibitemOpen
  \bibfield  {author} {\bibinfo {author} {\bibfnamefont {S.~K.}\ \bibnamefont
  {Behura}}, \bibinfo {author} {\bibfnamefont {A.}~\bibnamefont {Miranda}},
  \bibinfo {author} {\bibfnamefont {S.}~\bibnamefont {Nayak}}, \bibinfo
  {author} {\bibfnamefont {K.}~\bibnamefont {Johnson}}, \bibinfo {author}
  {\bibfnamefont {P.}~\bibnamefont {Das}},\ and\ \bibinfo {author}
  {\bibfnamefont {N.~R.}\ \bibnamefont {Pradhan}},\ }\bibfield  {title}
  {\bibinfo {title} {Moiré physics in twisted van der {Waals} heterostructures
  of {2D} materials},\ }\href@noop {} {\bibfield  {journal} {\bibinfo
  {journal} {Emergent Mater.}\ }\textbf {\bibinfo {volume} {4}},\ \bibinfo
  {pages} {813} (\bibinfo {year} {2021})}\BibitemShut {NoStop}%
\bibitem [{\citenamefont {Chen}\ \emph {et~al.}(2019)\citenamefont {Chen},
  \citenamefont {Jiang}, \citenamefont {Wu}, \citenamefont {Lyu}, \citenamefont
  {Li}, \citenamefont {Chittari}, \citenamefont {Watanabe}, \citenamefont
  {Taniguchi}, \citenamefont {Shi}, \citenamefont {Jung}, \citenamefont
  {Zhang},\ and\ \citenamefont {Wang}}]{chen_evidence_2019}%
  \BibitemOpen
  \bibfield  {author} {\bibinfo {author} {\bibfnamefont {G.}~\bibnamefont
  {Chen}}, \bibinfo {author} {\bibfnamefont {L.}~\bibnamefont {Jiang}},
  \bibinfo {author} {\bibfnamefont {S.}~\bibnamefont {Wu}}, \bibinfo {author}
  {\bibfnamefont {B.}~\bibnamefont {Lyu}}, \bibinfo {author} {\bibfnamefont
  {H.}~\bibnamefont {Li}}, \bibinfo {author} {\bibfnamefont {B.~L.}\
  \bibnamefont {Chittari}}, \bibinfo {author} {\bibfnamefont {K.}~\bibnamefont
  {Watanabe}}, \bibinfo {author} {\bibfnamefont {T.}~\bibnamefont {Taniguchi}},
  \bibinfo {author} {\bibfnamefont {Z.}~\bibnamefont {Shi}}, \bibinfo {author}
  {\bibfnamefont {J.}~\bibnamefont {Jung}}, \bibinfo {author} {\bibfnamefont
  {Y.}~\bibnamefont {Zhang}},\ and\ \bibinfo {author} {\bibfnamefont
  {F.}~\bibnamefont {Wang}},\ }\bibfield  {title} {\bibinfo {title} {Evidence
  of a gate-tunable {Mott} insulator in a trilayer graphene moiré
  superlattice},\ }\href {https://doi.org/10.1038/s41567-018-0387-2} {\bibfield
   {journal} {\bibinfo  {journal} {Nat. Phys.}\ }\textbf {\bibinfo {volume}
  {15}},\ \bibinfo {pages} {237} (\bibinfo {year} {2019})}\BibitemShut
  {NoStop}%
\bibitem [{\citenamefont {Cazalilla}\ \emph {et~al.}(2014)\citenamefont
  {Cazalilla}, \citenamefont {Ochoa},\ and\ \citenamefont
  {Guinea}}]{cazalilla_quantum_2014}%
  \BibitemOpen
  \bibfield  {author} {\bibinfo {author} {\bibfnamefont {M.}~\bibnamefont
  {Cazalilla}}, \bibinfo {author} {\bibfnamefont {H.}~\bibnamefont {Ochoa}},\
  and\ \bibinfo {author} {\bibfnamefont {F.}~\bibnamefont {Guinea}},\
  }\bibfield  {title} {\bibinfo {title} {Quantum {Spin} {Hall} {Effect} in
  {Two}-{Dimensional} {Crystals} of {Transition}-{Metal} {Dichalcogenides}},\
  }\href {https://doi.org/10.1103/PhysRevLett.113.077201} {\bibfield  {journal}
  {\bibinfo  {journal} {Phys. Rev. Lett.}\ }\textbf {\bibinfo {volume} {113}},\
  \bibinfo {pages} {077201} (\bibinfo {year} {2014})}\BibitemShut {NoStop}%
\bibitem [{\citenamefont {Ghorashi}\ \emph {et~al.}(2023)\citenamefont
  {Ghorashi}, \citenamefont {Dunbrack}, \citenamefont {Abouelkomsan},
  \citenamefont {Sun}, \citenamefont {Du},\ and\ \citenamefont
  {Cano}}]{ghorashi_topological_2023}%
  \BibitemOpen
  \bibfield  {author} {\bibinfo {author} {\bibfnamefont {S.~A.~A.}\
  \bibnamefont {Ghorashi}}, \bibinfo {author} {\bibfnamefont {A.}~\bibnamefont
  {Dunbrack}}, \bibinfo {author} {\bibfnamefont {A.}~\bibnamefont
  {Abouelkomsan}}, \bibinfo {author} {\bibfnamefont {J.}~\bibnamefont {Sun}},
  \bibinfo {author} {\bibfnamefont {X.}~\bibnamefont {Du}},\ and\ \bibinfo
  {author} {\bibfnamefont {J.}~\bibnamefont {Cano}},\ }\bibfield  {title}
  {\bibinfo {title} {Topological and {Stacked} {Flat} {Bands} in {Bilayer}
  {Graphene} with a {Superlattice} {Potential}},\ }\href
  {https://doi.org/10.1103/PhysRevLett.130.196201} {\bibfield  {journal}
  {\bibinfo  {journal} {Phys. Rev. Lett.}\ }\textbf {\bibinfo {volume} {130}},\
  \bibinfo {pages} {196201} (\bibinfo {year} {2023})}\BibitemShut {NoStop}%
\bibitem [{\citenamefont {He}\ \emph {et~al.}(2021)\citenamefont {He},
  \citenamefont {Zhou}, \citenamefont {Ye}, \citenamefont {Cho}, \citenamefont
  {Jeong}, \citenamefont {Meng},\ and\ \citenamefont {Wang}}]{he_moire_2021}%
  \BibitemOpen
  \bibfield  {author} {\bibinfo {author} {\bibfnamefont {F.}~\bibnamefont
  {He}}, \bibinfo {author} {\bibfnamefont {Y.}~\bibnamefont {Zhou}}, \bibinfo
  {author} {\bibfnamefont {Z.}~\bibnamefont {Ye}}, \bibinfo {author}
  {\bibfnamefont {S.-H.}\ \bibnamefont {Cho}}, \bibinfo {author} {\bibfnamefont
  {J.}~\bibnamefont {Jeong}}, \bibinfo {author} {\bibfnamefont
  {X.}~\bibnamefont {Meng}},\ and\ \bibinfo {author} {\bibfnamefont
  {Y.}~\bibnamefont {Wang}},\ }\bibfield  {title} {\bibinfo {title} {Moir{\'e}
  {Patterns} in {2D} {Materials}: {A} {Review}},\ }\href
  {https://doi.org/10.1021/acsnano.0c10435} {\bibfield  {journal} {\bibinfo
  {journal} {ACS Nano}\ }\textbf {\bibinfo {volume} {15}},\ \bibinfo {pages}
  {5944} (\bibinfo {year} {2021})}\BibitemShut {NoStop}%
\bibitem [{\citenamefont {Kennes}\ \emph {et~al.}(2021)\citenamefont {Kennes},
  \citenamefont {Claassen}, \citenamefont {Xian}, \citenamefont {Georges},
  \citenamefont {Millis}, \citenamefont {Hone}, \citenamefont {Dean},
  \citenamefont {Basov}, \citenamefont {Pasupathy},\ and\ \citenamefont
  {Rubio}}]{kennes_moire_2021}%
  \BibitemOpen
  \bibfield  {author} {\bibinfo {author} {\bibfnamefont {D.~M.}\ \bibnamefont
  {Kennes}}, \bibinfo {author} {\bibfnamefont {M.}~\bibnamefont {Claassen}},
  \bibinfo {author} {\bibfnamefont {L.}~\bibnamefont {Xian}}, \bibinfo {author}
  {\bibfnamefont {A.}~\bibnamefont {Georges}}, \bibinfo {author} {\bibfnamefont
  {A.~J.}\ \bibnamefont {Millis}}, \bibinfo {author} {\bibfnamefont
  {J.}~\bibnamefont {Hone}}, \bibinfo {author} {\bibfnamefont {C.~R.}\
  \bibnamefont {Dean}}, \bibinfo {author} {\bibfnamefont {D.~N.}\ \bibnamefont
  {Basov}}, \bibinfo {author} {\bibfnamefont {A.~N.}\ \bibnamefont
  {Pasupathy}},\ and\ \bibinfo {author} {\bibfnamefont {A.}~\bibnamefont
  {Rubio}},\ }\bibfield  {title} {\bibinfo {title} {Moiré heterostructures as
  a condensed-matter quantum simulator},\ }\href@noop {} {\bibfield  {journal}
  {\bibinfo  {journal} {Nat. Phys.}\ }\textbf {\bibinfo {volume} {17}},\
  \bibinfo {pages} {155} (\bibinfo {year} {2021})}\BibitemShut {NoStop}%
\bibitem [{\citenamefont {Kögl}\ \emph {et~al.}(2023)\citenamefont {Kögl},
  \citenamefont {Soubelet}, \citenamefont {Brotons-Gisbert}, \citenamefont
  {Stier}, \citenamefont {Gerardot},\ and\ \citenamefont
  {Finley}}]{kogl_moire_2023}%
  \BibitemOpen
  \bibfield  {author} {\bibinfo {author} {\bibfnamefont {M.}~\bibnamefont
  {Kögl}}, \bibinfo {author} {\bibfnamefont {P.}~\bibnamefont {Soubelet}},
  \bibinfo {author} {\bibfnamefont {M.}~\bibnamefont {Brotons-Gisbert}},
  \bibinfo {author} {\bibfnamefont {A.~V.}\ \bibnamefont {Stier}}, \bibinfo
  {author} {\bibfnamefont {B.~D.}\ \bibnamefont {Gerardot}},\ and\ \bibinfo
  {author} {\bibfnamefont {J.~J.}\ \bibnamefont {Finley}},\ }\bibfield  {title}
  {\bibinfo {title} {Moir{\'e} straintronics: a universal platform for
  reconfigurable quantum materials},\ }\href@noop {} {\bibfield  {journal}
  {\bibinfo  {journal} {npj 2D Mater Appl}\ }\textbf {\bibinfo {volume} {7}},\
  \bibinfo {pages} {1} (\bibinfo {year} {2023})}\BibitemShut {NoStop}%
\bibitem [{\citenamefont {Dean}\ \emph {et~al.}(2013)\citenamefont {Dean},
  \citenamefont {Wang}, \citenamefont {Maher}, \citenamefont {Forsythe},
  \citenamefont {Ghahari}, \citenamefont {Gao}, \citenamefont {Katoch},
  \citenamefont {Ishigami}, \citenamefont {Moon}, \citenamefont {Koshino},
  \citenamefont {Taniguchi}, \citenamefont {Watanabe}, \citenamefont {Shepard},
  \citenamefont {Hone},\ and\ \citenamefont {Kim}}]{dean_hofstadters_2013}%
  \BibitemOpen
  \bibfield  {author} {\bibinfo {author} {\bibfnamefont {C.~R.}\ \bibnamefont
  {Dean}}, \bibinfo {author} {\bibfnamefont {L.}~\bibnamefont {Wang}}, \bibinfo
  {author} {\bibfnamefont {P.}~\bibnamefont {Maher}}, \bibinfo {author}
  {\bibfnamefont {C.}~\bibnamefont {Forsythe}}, \bibinfo {author}
  {\bibfnamefont {F.}~\bibnamefont {Ghahari}}, \bibinfo {author} {\bibfnamefont
  {Y.}~\bibnamefont {Gao}}, \bibinfo {author} {\bibfnamefont {J.}~\bibnamefont
  {Katoch}}, \bibinfo {author} {\bibfnamefont {M.}~\bibnamefont {Ishigami}},
  \bibinfo {author} {\bibfnamefont {P.}~\bibnamefont {Moon}}, \bibinfo {author}
  {\bibfnamefont {M.}~\bibnamefont {Koshino}}, \bibinfo {author} {\bibfnamefont
  {T.}~\bibnamefont {Taniguchi}}, \bibinfo {author} {\bibfnamefont
  {K.}~\bibnamefont {Watanabe}}, \bibinfo {author} {\bibfnamefont {K.~L.}\
  \bibnamefont {Shepard}}, \bibinfo {author} {\bibfnamefont {J.}~\bibnamefont
  {Hone}},\ and\ \bibinfo {author} {\bibfnamefont {P.}~\bibnamefont {Kim}},\
  }\bibfield  {title} {\bibinfo {title} {Hofstadter’s butterfly and the
  fractal quantum {Hall} effect in moiré superlattices},\ }\href@noop {}
  {\bibfield  {journal} {\bibinfo  {journal} {Nature}\ }\textbf {\bibinfo
  {volume} {497}},\ \bibinfo {pages} {598} (\bibinfo {year}
  {2013})}\BibitemShut {NoStop}%
\bibitem [{\citenamefont {Li}\ \emph {et~al.}(2015)\citenamefont {Li},
  \citenamefont {Contryman}, \citenamefont {Qian}, \citenamefont {Ardakani},
  \citenamefont {Gong}, \citenamefont {Wang}, \citenamefont {Weisse},
  \citenamefont {Lee}, \citenamefont {Zhao}, \citenamefont {Ajayan},
  \citenamefont {Li}, \citenamefont {Manoharan},\ and\ \citenamefont
  {Zheng}}]{li_optoelectronic_2015}%
  \BibitemOpen
  \bibfield  {author} {\bibinfo {author} {\bibfnamefont {H.}~\bibnamefont
  {Li}}, \bibinfo {author} {\bibfnamefont {A.~W.}\ \bibnamefont {Contryman}},
  \bibinfo {author} {\bibfnamefont {X.}~\bibnamefont {Qian}}, \bibinfo {author}
  {\bibfnamefont {S.~M.}\ \bibnamefont {Ardakani}}, \bibinfo {author}
  {\bibfnamefont {Y.}~\bibnamefont {Gong}}, \bibinfo {author} {\bibfnamefont
  {X.}~\bibnamefont {Wang}}, \bibinfo {author} {\bibfnamefont {J.~M.}\
  \bibnamefont {Weisse}}, \bibinfo {author} {\bibfnamefont {C.~H.}\
  \bibnamefont {Lee}}, \bibinfo {author} {\bibfnamefont {J.}~\bibnamefont
  {Zhao}}, \bibinfo {author} {\bibfnamefont {P.~M.}\ \bibnamefont {Ajayan}},
  \bibinfo {author} {\bibfnamefont {J.}~\bibnamefont {Li}}, \bibinfo {author}
  {\bibfnamefont {H.~C.}\ \bibnamefont {Manoharan}},\ and\ \bibinfo {author}
  {\bibfnamefont {X.}~\bibnamefont {Zheng}},\ }\bibfield  {title} {\bibinfo
  {title} {Optoelectronic crystal of artificial atoms in strain-textured
  molybdenum disulphide},\ }\href@noop {} {\bibfield  {journal} {\bibinfo
  {journal} {Nat. Commun.}\ }\textbf {\bibinfo {volume} {6}},\ \bibinfo {pages}
  {1} (\bibinfo {year} {2015})}\BibitemShut {NoStop}%
\bibitem [{\citenamefont {Banerjee}\ \emph {et~al.}(2020)\citenamefont
  {Banerjee}, \citenamefont {Nguyen}, \citenamefont {Granzier-Nakajima},
  \citenamefont {Pabbi}, \citenamefont {Lherbier}, \citenamefont {Binion},
  \citenamefont {Charlier}, \citenamefont {Terrones},\ and\ \citenamefont
  {Hudson}}]{banerjee_strain_2020}%
  \BibitemOpen
  \bibfield  {author} {\bibinfo {author} {\bibfnamefont {R.}~\bibnamefont
  {Banerjee}}, \bibinfo {author} {\bibfnamefont {V.-H.}\ \bibnamefont
  {Nguyen}}, \bibinfo {author} {\bibfnamefont {T.}~\bibnamefont
  {Granzier-Nakajima}}, \bibinfo {author} {\bibfnamefont {L.}~\bibnamefont
  {Pabbi}}, \bibinfo {author} {\bibfnamefont {A.}~\bibnamefont {Lherbier}},
  \bibinfo {author} {\bibfnamefont {A.~R.}\ \bibnamefont {Binion}}, \bibinfo
  {author} {\bibfnamefont {J.-C.}\ \bibnamefont {Charlier}}, \bibinfo {author}
  {\bibfnamefont {M.}~\bibnamefont {Terrones}},\ and\ \bibinfo {author}
  {\bibfnamefont {E.~W.}\ \bibnamefont {Hudson}},\ }\bibfield  {title}
  {\bibinfo {title} {Strain {Modulated} {Superlattices} in {Graphene}},\
  }\href@noop {} {\bibfield  {journal} {\bibinfo  {journal} {Nano Lett.}\
  }\textbf {\bibinfo {volume} {20}},\ \bibinfo {pages} {3113} (\bibinfo {year}
  {2020})}\BibitemShut {NoStop}%
\bibitem [{\citenamefont {Zhang}\ \emph {et~al.}(2019)\citenamefont {Zhang},
  \citenamefont {Kim}, \citenamefont {Gilbert},\ and\ \citenamefont
  {Mason}}]{zhang_magnetotransport_2019}%
  \BibitemOpen
  \bibfield  {author} {\bibinfo {author} {\bibfnamefont {Y.}~\bibnamefont
  {Zhang}}, \bibinfo {author} {\bibfnamefont {Y.}~\bibnamefont {Kim}}, \bibinfo
  {author} {\bibfnamefont {M.~J.}\ \bibnamefont {Gilbert}},\ and\ \bibinfo
  {author} {\bibfnamefont {N.}~\bibnamefont {Mason}},\ }\bibfield  {title}
  {\bibinfo {title} {Magnetotransport in a strain superlattice of graphene},\
  }\href@noop {} {\bibfield  {journal} {\bibinfo  {journal} {Applied Physics
  Letters}\ }\textbf {\bibinfo {volume} {115}},\ \bibinfo {pages} {143508}
  (\bibinfo {year} {2019})}\BibitemShut {NoStop}%
\bibitem [{\citenamefont {Park}\ \emph {et~al.}(2008)\citenamefont {Park},
  \citenamefont {Son}, \citenamefont {Yang}, \citenamefont {Cohen},\ and\
  \citenamefont {Louie}}]{park_electron_2008}%
  \BibitemOpen
  \bibfield  {author} {\bibinfo {author} {\bibfnamefont {C.-H.}\ \bibnamefont
  {Park}}, \bibinfo {author} {\bibfnamefont {Y.-W.}\ \bibnamefont {Son}},
  \bibinfo {author} {\bibfnamefont {L.}~\bibnamefont {Yang}}, \bibinfo {author}
  {\bibfnamefont {M.~L.}\ \bibnamefont {Cohen}},\ and\ \bibinfo {author}
  {\bibfnamefont {S.~G.}\ \bibnamefont {Louie}},\ }\bibfield  {title} {\bibinfo
  {title} {Electron {Beam} {Supercollimation} in {Graphene} {Superlattices}},\
  }\href@noop {} {\bibfield  {journal} {\bibinfo  {journal} {Nano Lett.}\
  }\textbf {\bibinfo {volume} {8}},\ \bibinfo {pages} {2920} (\bibinfo {year}
  {2008})}\BibitemShut {NoStop}%
\bibitem [{\citenamefont {Lin}\ and\ \citenamefont
  {Tom{\'a}nek}(2020)}]{lin_periodically_2020}%
  \BibitemOpen
  \bibfield  {author} {\bibinfo {author} {\bibfnamefont {X.}~\bibnamefont
  {Lin}}\ and\ \bibinfo {author} {\bibfnamefont {D.}~\bibnamefont
  {Tom{\'a}nek}},\ }\bibfield  {title} {\bibinfo {title} {Periodically {Gated}
  {Bilayer} {Graphene} as an {Electronic} {Metamaterial}},\ }\href@noop {}
  {\bibfield  {journal} {\bibinfo  {journal} {Phys. Rev. Appl.}\ }\textbf
  {\bibinfo {volume} {13}},\ \bibinfo {pages} {034034} (\bibinfo {year}
  {2020})}\BibitemShut {NoStop}%
\bibitem [{\citenamefont {Hui-yun}\ \emph {et~al.}(2011)\citenamefont
  {Hui-yun}, \citenamefont {Ying}, \citenamefont {Yu-ping}, \citenamefont
  {Shi-lin},\ and\ \citenamefont {Shi-fan}}]{hui-yun_tunable_2011}%
  \BibitemOpen
  \bibfield  {author} {\bibinfo {author} {\bibfnamefont {Z.}~\bibnamefont
  {Hui-yun}}, \bibinfo {author} {\bibfnamefont {G.}~\bibnamefont {Ying}},
  \bibinfo {author} {\bibfnamefont {Z.}~\bibnamefont {Yu-ping}}, \bibinfo
  {author} {\bibfnamefont {X.}~\bibnamefont {Shi-lin}},\ and\ \bibinfo {author}
  {\bibfnamefont {W.}~\bibnamefont {Shi-fan}},\ }\bibfield  {title} {\bibinfo
  {title} {A tunable electron wave filter based on graphene superlattices with
  periodic potential patterns},\ }\href@noop {} {\bibfield  {journal} {\bibinfo
   {journal} {Appl. Phys. Lett.}\ }\textbf {\bibinfo {volume} {99}},\ \bibinfo
  {pages} {1} (\bibinfo {year} {2011})}\BibitemShut {NoStop}%
\bibitem [{\citenamefont {Chen}\ \emph {et~al.}(2020)\citenamefont {Chen},
  \citenamefont {Kraft}, \citenamefont {Danneau}, \citenamefont {Richter},\
  and\ \citenamefont {Liu}}]{chen_electrostatic_2020}%
  \BibitemOpen
  \bibfield  {author} {\bibinfo {author} {\bibfnamefont {S.-C.}\ \bibnamefont
  {Chen}}, \bibinfo {author} {\bibfnamefont {R.}~\bibnamefont {Kraft}},
  \bibinfo {author} {\bibfnamefont {R.}~\bibnamefont {Danneau}}, \bibinfo
  {author} {\bibfnamefont {K.}~\bibnamefont {Richter}},\ and\ \bibinfo {author}
  {\bibfnamefont {M.-H.}\ \bibnamefont {Liu}},\ }\bibfield  {title} {\bibinfo
  {title} {Electrostatic superlattices on scaled graphene lattices},\
  }\href@noop {} {\bibfield  {journal} {\bibinfo  {journal} {Communications
  Physics}\ }\textbf {\bibinfo {volume} {3}},\ \bibinfo {pages} {1} (\bibinfo
  {year} {2020})}\BibitemShut {NoStop}%
\bibitem [{\citenamefont {Brey}\ and\ \citenamefont
  {Fertig}(2009)}]{brey_emerging_2009}%
  \BibitemOpen
  \bibfield  {author} {\bibinfo {author} {\bibfnamefont {L.}~\bibnamefont
  {Brey}}\ and\ \bibinfo {author} {\bibfnamefont {H.~A.}\ \bibnamefont
  {Fertig}},\ }\bibfield  {title} {\bibinfo {title} {Emerging {Zero} {Modes}
  for {Graphene} in a {Periodic} {Potential}},\ }\href@noop {} {\bibfield
  {journal} {\bibinfo  {journal} {Phys. Rev. Lett.}\ }\textbf {\bibinfo
  {volume} {103}},\ \bibinfo {pages} {046809} (\bibinfo {year}
  {2009})}\BibitemShut {NoStop}%
\bibitem [{\citenamefont {Wu}\ \emph {et~al.}(2012)\citenamefont {Wu},
  \citenamefont {Killi},\ and\ \citenamefont {Paramekanti}}]{wu_graphene_2012}%
  \BibitemOpen
  \bibfield  {author} {\bibinfo {author} {\bibfnamefont {S.}~\bibnamefont
  {Wu}}, \bibinfo {author} {\bibfnamefont {M.}~\bibnamefont {Killi}},\ and\
  \bibinfo {author} {\bibfnamefont {A.}~\bibnamefont {Paramekanti}},\
  }\bibfield  {title} {\bibinfo {title} {Graphene under spatially varying
  external potentials: {Landau} levels, magnetotransport, and topological
  modes},\ }\href@noop {} {\bibfield  {journal} {\bibinfo  {journal} {Phys.
  Rev. B}\ }\textbf {\bibinfo {volume} {85}},\ \bibinfo {pages} {195404}
  (\bibinfo {year} {2012})}\BibitemShut {NoStop}%
\bibitem [{\citenamefont {Ono}(2017)}]{ono_effect_2017}%
  \BibitemOpen
  \bibfield  {author} {\bibinfo {author} {\bibfnamefont {S.}~\bibnamefont
  {Ono}},\ }\bibfield  {title} {\bibinfo {title} {Effect of one-dimensional
  superlattice potentials on the band gap of two-dimensional materials},\
  }\href {https://doi.org/10.1063/1.4984069} {\bibfield  {journal} {\bibinfo
  {journal} {J. Appl. Phys.}\ }\textbf {\bibinfo {volume} {121}},\ \bibinfo
  {pages} {204301} (\bibinfo {year} {2017})}\BibitemShut {NoStop}%
\bibitem [{\citenamefont {Sattari}\ and\ \citenamefont
  {Mirershadi}(2021)}]{sattari_effect_2021}%
  \BibitemOpen
  \bibfield  {author} {\bibinfo {author} {\bibfnamefont {F.}~\bibnamefont
  {Sattari}}\ and\ \bibinfo {author} {\bibfnamefont {S.}~\bibnamefont
  {Mirershadi}},\ }\bibfield  {title} {\bibinfo {title} {Effect of the strain
  on spin-valley transport properties in {MoS$_2$} superlattice},\ }\href@noop
  {} {\bibfield  {journal} {\bibinfo  {journal} {Sci. Rep.}\ }\textbf {\bibinfo
  {volume} {11}},\ \bibinfo {pages} {17617} (\bibinfo {year}
  {2021})}\BibitemShut {NoStop}%
\bibitem [{\citenamefont {Chaves}\ \emph {et~al.}(2020)\citenamefont {Chaves},
  \citenamefont {Azadani}, \citenamefont {Alsalman}, \citenamefont {da~Costa},
  \citenamefont {Frisenda}, \citenamefont {Chaves}, \citenamefont {Song},
  \citenamefont {Kim}, \citenamefont {He}, \citenamefont {Zhou}, \citenamefont
  {Castellanos-Gomez}, \citenamefont {Peeters}, \citenamefont {Liu},
  \citenamefont {Hinkle}, \citenamefont {Oh}, \citenamefont {Ye}, \citenamefont
  {Koester}, \citenamefont {Lee}, \citenamefont {Avouris}, \citenamefont
  {Wang},\ and\ \citenamefont {Low}}]{chaves_bandgap_2020}%
  \BibitemOpen
  \bibfield  {author} {\bibinfo {author} {\bibfnamefont {A.}~\bibnamefont
  {Chaves}}, \bibinfo {author} {\bibfnamefont {J.~G.}\ \bibnamefont {Azadani}},
  \bibinfo {author} {\bibfnamefont {H.}~\bibnamefont {Alsalman}}, \bibinfo
  {author} {\bibfnamefont {D.~R.}\ \bibnamefont {da~Costa}}, \bibinfo {author}
  {\bibfnamefont {R.}~\bibnamefont {Frisenda}}, \bibinfo {author}
  {\bibfnamefont {A.~J.}\ \bibnamefont {Chaves}}, \bibinfo {author}
  {\bibfnamefont {S.~H.}\ \bibnamefont {Song}}, \bibinfo {author}
  {\bibfnamefont {Y.~D.}\ \bibnamefont {Kim}}, \bibinfo {author} {\bibfnamefont
  {D.}~\bibnamefont {He}}, \bibinfo {author} {\bibfnamefont {J.}~\bibnamefont
  {Zhou}}, \bibinfo {author} {\bibfnamefont {A.}~\bibnamefont
  {Castellanos-Gomez}}, \bibinfo {author} {\bibfnamefont {F.~M.}\ \bibnamefont
  {Peeters}}, \bibinfo {author} {\bibfnamefont {Z.}~\bibnamefont {Liu}},
  \bibinfo {author} {\bibfnamefont {C.~L.}\ \bibnamefont {Hinkle}}, \bibinfo
  {author} {\bibfnamefont {S.-H.}\ \bibnamefont {Oh}}, \bibinfo {author}
  {\bibfnamefont {P.~D.}\ \bibnamefont {Ye}}, \bibinfo {author} {\bibfnamefont
  {S.~J.}\ \bibnamefont {Koester}}, \bibinfo {author} {\bibfnamefont {Y.~H.}\
  \bibnamefont {Lee}}, \bibinfo {author} {\bibfnamefont {P.}~\bibnamefont
  {Avouris}}, \bibinfo {author} {\bibfnamefont {X.}~\bibnamefont {Wang}},\ and\
  \bibinfo {author} {\bibfnamefont {T.}~\bibnamefont {Low}},\ }\bibfield
  {title} {\bibinfo {title} {Bandgap engineering of two-dimensional
  semiconductor materials},\ }\href
  {https://doi.org/10.1038/s41699-020-00162-4} {\bibfield  {journal} {\bibinfo
  {journal} {npj 2D Mater Appl}\ }\textbf {\bibinfo {volume} {4}},\ \bibinfo
  {pages} {1} (\bibinfo {year} {2020})}\BibitemShut {NoStop}%
\bibitem [{\citenamefont {Conley}\ \emph {et~al.}(2013)\citenamefont {Conley},
  \citenamefont {Wang}, \citenamefont {Ziegler}, \citenamefont {Haglund},
  \citenamefont {Pantelides},\ and\ \citenamefont
  {Bolotin}}]{conley_bandgap_2013}%
  \BibitemOpen
  \bibfield  {author} {\bibinfo {author} {\bibfnamefont {H.~J.}\ \bibnamefont
  {Conley}}, \bibinfo {author} {\bibfnamefont {B.}~\bibnamefont {Wang}},
  \bibinfo {author} {\bibfnamefont {J.~I.}\ \bibnamefont {Ziegler}}, \bibinfo
  {author} {\bibfnamefont {R.~F.~J.}\ \bibnamefont {Haglund}}, \bibinfo
  {author} {\bibfnamefont {S.~T.}\ \bibnamefont {Pantelides}},\ and\ \bibinfo
  {author} {\bibfnamefont {K.~I.}\ \bibnamefont {Bolotin}},\ }\bibfield
  {title} {\bibinfo {title} {Bandgap {Engineering} of {Strained} {Monolayer}
  and {Bilayer} {MoS$_2$}},\ }\href {https://doi.org/10.1021/nl4014748}
  {\bibfield  {journal} {\bibinfo  {journal} {Nano Lett.}\ }\textbf {\bibinfo
  {volume} {13}},\ \bibinfo {pages} {3626} (\bibinfo {year}
  {2013})}\BibitemShut {NoStop}%
\bibitem [{\citenamefont {Manzeli}\ \emph {et~al.}(2015)\citenamefont
  {Manzeli}, \citenamefont {Allain}, \citenamefont {Ghadimi},\ and\
  \citenamefont {Kis}}]{manzeli_piezoresistivity_2015}%
  \BibitemOpen
  \bibfield  {author} {\bibinfo {author} {\bibfnamefont {S.}~\bibnamefont
  {Manzeli}}, \bibinfo {author} {\bibfnamefont {A.}~\bibnamefont {Allain}},
  \bibinfo {author} {\bibfnamefont {A.}~\bibnamefont {Ghadimi}},\ and\ \bibinfo
  {author} {\bibfnamefont {A.}~\bibnamefont {Kis}},\ }\bibfield  {title}
  {\bibinfo {title} {Piezoresistivity and strain-induced band gap tuning in
  atomically thin {MoS$_2$}},\ }\href
  {https://doi.org/10.1021/acs.nanolett.5b01689} {\bibfield  {journal}
  {\bibinfo  {journal} {Nano Lett.}\ }\textbf {\bibinfo {volume} {15}},\
  \bibinfo {pages} {5330} (\bibinfo {year} {2015})}\BibitemShut {NoStop}%
\bibitem [{\citenamefont {Yun}\ \emph {et~al.}(2012)\citenamefont {Yun},
  \citenamefont {Han}, \citenamefont {Hong}, \citenamefont {Kim},\ and\
  \citenamefont {Lee}}]{yun_thickness_2012}%
  \BibitemOpen
  \bibfield  {author} {\bibinfo {author} {\bibfnamefont {W.~S.}\ \bibnamefont
  {Yun}}, \bibinfo {author} {\bibfnamefont {S.~W.}\ \bibnamefont {Han}},
  \bibinfo {author} {\bibfnamefont {S.~C.}\ \bibnamefont {Hong}}, \bibinfo
  {author} {\bibfnamefont {I.~G.}\ \bibnamefont {Kim}},\ and\ \bibinfo {author}
  {\bibfnamefont {J.~D.}\ \bibnamefont {Lee}},\ }\bibfield  {title} {\bibinfo
  {title} {Thickness and strain effects on electronic structures of transition
  metal dichalcogenides: {2H}-{MX$_2$} semiconductors ({M=Mo, W; X=S, Se,
  Te})},\ }\href@noop {} {\bibfield  {journal} {\bibinfo  {journal} {Phys. Rev.
  B}\ }\textbf {\bibinfo {volume} {85}},\ \bibinfo {pages} {033305} (\bibinfo
  {year} {2012})}\BibitemShut {NoStop}%
\bibitem [{\citenamefont {Feng}\ \emph {et~al.}(2012)\citenamefont {Feng},
  \citenamefont {Qian}, \citenamefont {Huang},\ and\ \citenamefont
  {Li}}]{feng_strain-engineered_2012}%
  \BibitemOpen
  \bibfield  {author} {\bibinfo {author} {\bibfnamefont {J.}~\bibnamefont
  {Feng}}, \bibinfo {author} {\bibfnamefont {X.}~\bibnamefont {Qian}}, \bibinfo
  {author} {\bibfnamefont {C.-W.}\ \bibnamefont {Huang}},\ and\ \bibinfo
  {author} {\bibfnamefont {J.}~\bibnamefont {Li}},\ }\bibfield  {title}
  {\bibinfo {title} {Strain-engineered artificial atom as a broad-spectrum
  solar energy funnel},\ }\href@noop {} {\bibfield  {journal} {\bibinfo
  {journal} {Nat. Photonics}\ }\textbf {\bibinfo {volume} {6}},\ \bibinfo
  {pages} {866} (\bibinfo {year} {2012})}\BibitemShut {NoStop}%
\bibitem [{\citenamefont {Lee}\ \emph {et~al.}(2021)\citenamefont {Lee},
  \citenamefont {Yun}, \citenamefont {Seo}, \citenamefont {Cho}, \citenamefont
  {Kim}, \citenamefont {An}, \citenamefont {Kang}, \citenamefont {Lee},\ and\
  \citenamefont {Kim}}]{lee_switchable_2021}%
  \BibitemOpen
  \bibfield  {author} {\bibinfo {author} {\bibfnamefont {J.}~\bibnamefont
  {Lee}}, \bibinfo {author} {\bibfnamefont {S.~J.}\ \bibnamefont {Yun}},
  \bibinfo {author} {\bibfnamefont {C.}~\bibnamefont {Seo}}, \bibinfo {author}
  {\bibfnamefont {K.}~\bibnamefont {Cho}}, \bibinfo {author} {\bibfnamefont
  {T.~S.}\ \bibnamefont {Kim}}, \bibinfo {author} {\bibfnamefont {G.~H.}\
  \bibnamefont {An}}, \bibinfo {author} {\bibfnamefont {K.}~\bibnamefont
  {Kang}}, \bibinfo {author} {\bibfnamefont {H.~S.}\ \bibnamefont {Lee}},\ and\
  \bibinfo {author} {\bibfnamefont {J.}~\bibnamefont {Kim}},\ }\bibfield
  {title} {\bibinfo {title} {Switchable, {Tunable}, and {Directable} {Exciton}
  {Funneling} in {Periodically} {Wrinkled} {WS$_2$}},\ }\href@noop {}
  {\bibfield  {journal} {\bibinfo  {journal} {Nano Lett.}\ }\textbf {\bibinfo
  {volume} {21}},\ \bibinfo {pages} {43} (\bibinfo {year} {2021})}\BibitemShut
  {NoStop}%
\bibitem [{\citenamefont {Lee}\ \emph {et~al.}(2022)\citenamefont {Lee},
  \citenamefont {Koo}, \citenamefont {Choi}, \citenamefont {Kumar},
  \citenamefont {Lee}, \citenamefont {Ji}, \citenamefont {Choi}, \citenamefont
  {Kang}, \citenamefont {Kim}, \citenamefont {Park}, \citenamefont {Choo},\
  and\ \citenamefont {Park}}]{lee_drift-dominant_2022}%
  \BibitemOpen
  \bibfield  {author} {\bibinfo {author} {\bibfnamefont {H.}~\bibnamefont
  {Lee}}, \bibinfo {author} {\bibfnamefont {Y.}~\bibnamefont {Koo}}, \bibinfo
  {author} {\bibfnamefont {J.}~\bibnamefont {Choi}}, \bibinfo {author}
  {\bibfnamefont {S.}~\bibnamefont {Kumar}}, \bibinfo {author} {\bibfnamefont
  {H.-T.}\ \bibnamefont {Lee}}, \bibinfo {author} {\bibfnamefont
  {G.}~\bibnamefont {Ji}}, \bibinfo {author} {\bibfnamefont {S.~H.}\
  \bibnamefont {Choi}}, \bibinfo {author} {\bibfnamefont {M.}~\bibnamefont
  {Kang}}, \bibinfo {author} {\bibfnamefont {K.~K.}\ \bibnamefont {Kim}},
  \bibinfo {author} {\bibfnamefont {H.-R.}\ \bibnamefont {Park}}, \bibinfo
  {author} {\bibfnamefont {H.}~\bibnamefont {Choo}},\ and\ \bibinfo {author}
  {\bibfnamefont {K.-D.}\ \bibnamefont {Park}},\ }\bibfield  {title} {\bibinfo
  {title} {Drift-dominant exciton funneling and trion conversion in {2D}
  semiconductors on the nanogap},\ }\href@noop {} {\bibfield  {journal}
  {\bibinfo  {journal} {Science Advances}\ }\textbf {\bibinfo {volume} {8}},\
  \bibinfo {pages} {eabm5236} (\bibinfo {year} {2022})}\BibitemShut {NoStop}%
\bibitem [{\citenamefont {Waters}\ \emph {et~al.}(2020)\citenamefont {Waters},
  \citenamefont {Nie}, \citenamefont {Lüpke}, \citenamefont {Pan},
  \citenamefont {Fölsch}, \citenamefont {Lin}, \citenamefont {Jariwala},
  \citenamefont {Zhang}, \citenamefont {Wang}, \citenamefont {Lv},
  \citenamefont {Cho}, \citenamefont {Xiao}, \citenamefont {Robinson},\ and\
  \citenamefont {Feenstra}}]{waters_flat_2020}%
  \BibitemOpen
  \bibfield  {author} {\bibinfo {author} {\bibfnamefont {D.}~\bibnamefont
  {Waters}}, \bibinfo {author} {\bibfnamefont {Y.}~\bibnamefont {Nie}},
  \bibinfo {author} {\bibfnamefont {F.}~\bibnamefont {Lüpke}}, \bibinfo
  {author} {\bibfnamefont {Y.}~\bibnamefont {Pan}}, \bibinfo {author}
  {\bibfnamefont {S.}~\bibnamefont {Fölsch}}, \bibinfo {author} {\bibfnamefont
  {Y.-C.}\ \bibnamefont {Lin}}, \bibinfo {author} {\bibfnamefont
  {B.}~\bibnamefont {Jariwala}}, \bibinfo {author} {\bibfnamefont
  {K.}~\bibnamefont {Zhang}}, \bibinfo {author} {\bibfnamefont
  {C.}~\bibnamefont {Wang}}, \bibinfo {author} {\bibfnamefont {H.}~\bibnamefont
  {Lv}}, \bibinfo {author} {\bibfnamefont {K.}~\bibnamefont {Cho}}, \bibinfo
  {author} {\bibfnamefont {D.}~\bibnamefont {Xiao}}, \bibinfo {author}
  {\bibfnamefont {J.~A.}\ \bibnamefont {Robinson}},\ and\ \bibinfo {author}
  {\bibfnamefont {R.~M.}\ \bibnamefont {Feenstra}},\ }\bibfield  {title}
  {\bibinfo {title} {Flat {Bands} and {Mechanical} {Deformation} {Effects} in
  the {Moir{\'e}} {Superlattice} of {MoS$_2$}-{WSe$_2$} {Heterobilayers}},\
  }\href {https://doi.org/10.1021/acsnano.0c03414} {\bibfield  {journal}
  {\bibinfo  {journal} {ACS Nano}\ }\textbf {\bibinfo {volume} {14}},\ \bibinfo
  {pages} {7564} (\bibinfo {year} {2020})}\BibitemShut {NoStop}%
\bibitem [{\citenamefont {Rodr{\'i}guez}\ \emph {et~al.}(2023)\citenamefont
  {Rodr{\'i}guez}, \citenamefont {Varillas}, \citenamefont {Haider},
  \citenamefont {Kalb{\'a}{\v{c}}},\ and\ \citenamefont
  {Frank}}]{rodriguez_complex_2023}%
  \BibitemOpen
  \bibfield  {author} {\bibinfo {author} {\bibfnamefont {{\'A}.}~\bibnamefont
  {Rodr{\'i}guez}}, \bibinfo {author} {\bibfnamefont {J.}~\bibnamefont
  {Varillas}}, \bibinfo {author} {\bibfnamefont {G.}~\bibnamefont {Haider}},
  \bibinfo {author} {\bibfnamefont {M.}~\bibnamefont {Kalb{\'a}{\v{c}}}},\ and\
  \bibinfo {author} {\bibfnamefont {O.}~\bibnamefont {Frank}},\ }\bibfield
  {title} {\bibinfo {title} {Complex {Strain} {Scapes} in {Reconstructed}
  {Transition}-{Metal} {Dichalcogenide} {Moir{\'e}} {Superlattices}},\ }\href
  {https://doi.org/10.1021/acsnano.3c00609} {\bibfield  {journal} {\bibinfo
  {journal} {ACS Nano}\ }\textbf {\bibinfo {volume} {17}},\ \bibinfo {pages}
  {7787} (\bibinfo {year} {2023})}\BibitemShut {NoStop}%
\bibitem [{\citenamefont {Quan}\ \emph {et~al.}(2021)\citenamefont {Quan},
  \citenamefont {Linhart}, \citenamefont {Lin}, \citenamefont {Lee},
  \citenamefont {Zhu}, \citenamefont {Wang}, \citenamefont {Hsu}, \citenamefont
  {Choi}, \citenamefont {Embley}, \citenamefont {Young}, \citenamefont
  {Taniguchi}, \citenamefont {Watanabe}, \citenamefont {Shih}, \citenamefont
  {Lai}, \citenamefont {MacDonald}, \citenamefont {Tan}, \citenamefont
  {Libisch},\ and\ \citenamefont {Li}}]{quan_phonon_2021}%
  \BibitemOpen
  \bibfield  {author} {\bibinfo {author} {\bibfnamefont {J.}~\bibnamefont
  {Quan}}, \bibinfo {author} {\bibfnamefont {L.}~\bibnamefont {Linhart}},
  \bibinfo {author} {\bibfnamefont {M.-L.}\ \bibnamefont {Lin}}, \bibinfo
  {author} {\bibfnamefont {D.}~\bibnamefont {Lee}}, \bibinfo {author}
  {\bibfnamefont {J.}~\bibnamefont {Zhu}}, \bibinfo {author} {\bibfnamefont
  {C.-Y.}\ \bibnamefont {Wang}}, \bibinfo {author} {\bibfnamefont {W.-T.}\
  \bibnamefont {Hsu}}, \bibinfo {author} {\bibfnamefont {J.}~\bibnamefont
  {Choi}}, \bibinfo {author} {\bibfnamefont {J.}~\bibnamefont {Embley}},
  \bibinfo {author} {\bibfnamefont {C.}~\bibnamefont {Young}}, \bibinfo
  {author} {\bibfnamefont {T.}~\bibnamefont {Taniguchi}}, \bibinfo {author}
  {\bibfnamefont {K.}~\bibnamefont {Watanabe}}, \bibinfo {author}
  {\bibfnamefont {C.-K.}\ \bibnamefont {Shih}}, \bibinfo {author}
  {\bibfnamefont {K.}~\bibnamefont {Lai}}, \bibinfo {author} {\bibfnamefont
  {A.~H.}\ \bibnamefont {MacDonald}}, \bibinfo {author} {\bibfnamefont {P.-H.}\
  \bibnamefont {Tan}}, \bibinfo {author} {\bibfnamefont {F.}~\bibnamefont
  {Libisch}},\ and\ \bibinfo {author} {\bibfnamefont {X.}~\bibnamefont {Li}},\
  }\bibfield  {title} {\bibinfo {title} {Phonon renormalization in
  reconstructed {MoS$_2$} moiré superlattices},\ }\href@noop {} {\bibfield
  {journal} {\bibinfo  {journal} {Nat. Mater.}\ }\textbf {\bibinfo {volume}
  {20}},\ \bibinfo {pages} {1100} (\bibinfo {year} {2021})}\BibitemShut
  {NoStop}%
\bibitem [{\citenamefont {Zhu}\ and\ \citenamefont
  {Johnson}(2018)}]{zhu_moire-templated_2018}%
  \BibitemOpen
  \bibfield  {author} {\bibinfo {author} {\bibfnamefont {S.}~\bibnamefont
  {Zhu}}\ and\ \bibinfo {author} {\bibfnamefont {H.~T.}\ \bibnamefont
  {Johnson}},\ }\bibfield  {title} {\bibinfo {title} {Moir{\'e}-templated
  strain patterning in transition-metal dichalcogenides and application in
  twisted bilayer {MoS$_2$}},\ }\href@noop {} {\bibfield  {journal} {\bibinfo
  {journal} {Nanoscale}\ }\textbf {\bibinfo {volume} {10}},\ \bibinfo {pages}
  {20689} (\bibinfo {year} {2018})}\BibitemShut {NoStop}%
\bibitem [{\citenamefont {Zhang}\ \emph {et~al.}(2017)\citenamefont {Zhang},
  \citenamefont {Chuu}, \citenamefont {Ren}, \citenamefont {Li}, \citenamefont
  {Li}, \citenamefont {Jin}, \citenamefont {Chou},\ and\ \citenamefont
  {Shih}}]{zhang_interlayer_2017}%
  \BibitemOpen
  \bibfield  {author} {\bibinfo {author} {\bibfnamefont {C.}~\bibnamefont
  {Zhang}}, \bibinfo {author} {\bibfnamefont {C.-P.}\ \bibnamefont {Chuu}},
  \bibinfo {author} {\bibfnamefont {X.}~\bibnamefont {Ren}}, \bibinfo {author}
  {\bibfnamefont {M.-Y.}\ \bibnamefont {Li}}, \bibinfo {author} {\bibfnamefont
  {L.-J.}\ \bibnamefont {Li}}, \bibinfo {author} {\bibfnamefont
  {C.}~\bibnamefont {Jin}}, \bibinfo {author} {\bibfnamefont {M.-Y.}\
  \bibnamefont {Chou}},\ and\ \bibinfo {author} {\bibfnamefont {C.-K.}\
  \bibnamefont {Shih}},\ }\bibfield  {title} {\bibinfo {title} {Interlayer
  couplings, {Moir{\'e}} patterns, and {2D} electronic superlattices in
  {MoS$_2$}/{WSe$_2$} hetero-bilayers},\ }\href@noop {} {\bibfield  {journal}
  {\bibinfo  {journal} {Sci. Adv.}\ }\textbf {\bibinfo {volume} {3}},\ \bibinfo
  {pages} {e1601459} (\bibinfo {year} {2017})}\BibitemShut {NoStop}%
\bibitem [{\citenamefont {Duerloo}\ \emph {et~al.}(2012)\citenamefont
  {Duerloo}, \citenamefont {Ong},\ and\ \citenamefont
  {Reed}}]{duerloo_intrinsic_2012}%
  \BibitemOpen
  \bibfield  {author} {\bibinfo {author} {\bibfnamefont {K.}~\bibnamefont
  {Duerloo}}, \bibinfo {author} {\bibfnamefont {M.~T.}\ \bibnamefont {Ong}},\
  and\ \bibinfo {author} {\bibfnamefont {E.~J.}\ \bibnamefont {Reed}},\
  }\bibfield  {title} {\bibinfo {title} {Intrinsic piezoelectricity in
  two-dimensional materials},\ }\href
  {https://doi.org/10.1021/jz3012436&iName=master.img-000.jpg&w=131&h=110}
  {\bibfield  {journal} {\bibinfo  {journal} {J. Phys. Chem. Lett.}\ }\textbf
  {\bibinfo {volume} {3}},\ \bibinfo {pages} {2871} (\bibinfo {year}
  {2012})}\BibitemShut {NoStop}%
\bibitem [{\citenamefont {Wu}\ \emph {et~al.}(2014)\citenamefont {Wu},
  \citenamefont {Wang}, \citenamefont {Li}, \citenamefont {Zhang},
  \citenamefont {Lin}, \citenamefont {Niu}, \citenamefont {Chenet},
  \citenamefont {Zhang}, \citenamefont {Hao}, \citenamefont {Heinz},
  \citenamefont {Hone},\ and\ \citenamefont {Wang}}]{wu_piezoelectricity_2014}%
  \BibitemOpen
  \bibfield  {author} {\bibinfo {author} {\bibfnamefont {W.}~\bibnamefont
  {Wu}}, \bibinfo {author} {\bibfnamefont {L.}~\bibnamefont {Wang}}, \bibinfo
  {author} {\bibfnamefont {Y.}~\bibnamefont {Li}}, \bibinfo {author}
  {\bibfnamefont {F.}~\bibnamefont {Zhang}}, \bibinfo {author} {\bibfnamefont
  {L.}~\bibnamefont {Lin}}, \bibinfo {author} {\bibfnamefont {S.}~\bibnamefont
  {Niu}}, \bibinfo {author} {\bibfnamefont {D.}~\bibnamefont {Chenet}},
  \bibinfo {author} {\bibfnamefont {X.}~\bibnamefont {Zhang}}, \bibinfo
  {author} {\bibfnamefont {Y.}~\bibnamefont {Hao}}, \bibinfo {author}
  {\bibfnamefont {T.~F.}\ \bibnamefont {Heinz}}, \bibinfo {author}
  {\bibfnamefont {J.}~\bibnamefont {Hone}},\ and\ \bibinfo {author}
  {\bibfnamefont {Z.~L.}\ \bibnamefont {Wang}},\ }\bibfield  {title} {\bibinfo
  {title} {Piezoelectricity of single-atomic-layer {MoS$_2$} for energy
  conversion and piezotronics},\ }\href@noop {} {\bibfield  {journal} {\bibinfo
   {journal} {Nature}\ }\textbf {\bibinfo {volume} {514}},\ \bibinfo {pages}
  {470} (\bibinfo {year} {2014})}\BibitemShut {NoStop}%
\bibitem [{\citenamefont {Xiao}\ \emph {et~al.}(2012)\citenamefont {Xiao},
  \citenamefont {Liu}, \citenamefont {Feng}, \citenamefont {Xu},\ and\
  \citenamefont {Yao}}]{xiao_coupled_2012}%
  \BibitemOpen
  \bibfield  {author} {\bibinfo {author} {\bibfnamefont {D.}~\bibnamefont
  {Xiao}}, \bibinfo {author} {\bibfnamefont {G.-B.}\ \bibnamefont {Liu}},
  \bibinfo {author} {\bibfnamefont {W.}~\bibnamefont {Feng}}, \bibinfo {author}
  {\bibfnamefont {X.}~\bibnamefont {Xu}},\ and\ \bibinfo {author}
  {\bibfnamefont {W.}~\bibnamefont {Yao}},\ }\bibfield  {title} {\bibinfo
  {title} {Coupled {Spin} and {Valley} {Physics} in {Monolayers} of {MoS$_2$}
  and {Other} {Group}-{VI} {Dichalcogenides}},\ }\href
  {https://doi.org/10.1103/PhysRevLett.108.196802} {\bibfield  {journal}
  {\bibinfo  {journal} {Phys. Rev. Lett.}\ }\textbf {\bibinfo {volume} {108}},\
  \bibinfo {pages} {196802} (\bibinfo {year} {2012})}\BibitemShut {NoStop}%
\bibitem [{\citenamefont {Mak}\ \emph {et~al.}(2012)\citenamefont {Mak},
  \citenamefont {He}, \citenamefont {Shan},\ and\ \citenamefont
  {Heinz}}]{mak_control_2012}%
  \BibitemOpen
  \bibfield  {author} {\bibinfo {author} {\bibfnamefont {K.~F.}\ \bibnamefont
  {Mak}}, \bibinfo {author} {\bibfnamefont {K.}~\bibnamefont {He}}, \bibinfo
  {author} {\bibfnamefont {J.}~\bibnamefont {Shan}},\ and\ \bibinfo {author}
  {\bibfnamefont {T.~F.}\ \bibnamefont {Heinz}},\ }\bibfield  {title} {\bibinfo
  {title} {Control of valley polarization in monolayer {MoS$_2$} by optical
  helicity},\ }\href {https://doi.org/10.1038/nnano.2012.96} {\bibfield
  {journal} {\bibinfo  {journal} {Nat. Nanotechnol.}\ }\textbf {\bibinfo
  {volume} {7}},\ \bibinfo {pages} {494} (\bibinfo {year} {2012})}\BibitemShut
  {NoStop}%
\bibitem [{\citenamefont {Nye}(1985)}]{nye_physical_1985}%
  \BibitemOpen
  \bibfield  {author} {\bibinfo {author} {\bibfnamefont {J.~F.}\ \bibnamefont
  {Nye}},\ }\href@noop {} {\emph {\bibinfo {title} {Physical properties of
  crystals: their representation by tensors and matrices}}}\ (\bibinfo
  {publisher} {Oxford university press, New York},\ \bibinfo {year} {1985})\BibitemShut
  {NoStop}%
\bibitem [{Note1()}]{Note1}%
  \BibitemOpen
  \bibinfo {note} {See Supplemental Material at [URL will be inserted by
  publisher] for or supplemental text on piezoelectric response of monolayer MoS$_{2}$, Figs. S1 to S10, Table S1, and Refs. [72, 73].}\BibitemShut {Stop}%
\bibitem [{Note2()}]{Note2}%
  \BibitemOpen
  \bibinfo {note} {The areal dimensions of the supercell are 176.41 \r A$\times
  $ 3.18 \r A~while the thickness of the monolayer is 3.13 \r A.}\BibitemShut
  {Stop}%
\bibitem [{\citenamefont {Horiuchi}\ and\ \citenamefont
  {Tokura}(2008)}]{horiuchi_organic_2008}%
  \BibitemOpen
  \bibfield  {author} {\bibinfo {author} {\bibfnamefont {S.}~\bibnamefont
  {Horiuchi}}\ and\ \bibinfo {author} {\bibfnamefont {Y.}~\bibnamefont
  {Tokura}},\ }\bibfield  {title} {\bibinfo {title} {Organic ferroelectrics},\
  }\href {https://doi.org/10.1038/nmat2137} {\bibfield  {journal} {\bibinfo
  {journal} {Nat. Mater.}\ }\textbf {\bibinfo {volume} {7}},\ \bibinfo {pages}
  {357} (\bibinfo {year} {2008})}\BibitemShut {NoStop}%
\bibitem [{\citenamefont {Ku}\ \emph {et~al.}(2010)\citenamefont {Ku},
  \citenamefont {Berlijn},\ and\ \citenamefont {Lee}}]{ku_unfolding_2010}%
  \BibitemOpen
  \bibfield  {author} {\bibinfo {author} {\bibfnamefont {W.}~\bibnamefont
  {Ku}}, \bibinfo {author} {\bibfnamefont {T.}~\bibnamefont {Berlijn}},\ and\
  \bibinfo {author} {\bibfnamefont {C.-C.}\ \bibnamefont {Lee}},\ }\bibfield
  {title} {\bibinfo {title} {Unfolding {First}-{Principles} {Band}
  {Structures}},\ }\href {https://doi.org/10.1103/PhysRevLett.104.216401}
  {\bibfield  {journal} {\bibinfo  {journal} {Phys. Rev. Lett.}\ }\textbf
  {\bibinfo {volume} {104}},\ \bibinfo {pages} {216401} (\bibinfo {year}
  {2010})}\BibitemShut {NoStop}%
\bibitem [{\citenamefont {Popescu}\ and\ \citenamefont
  {Zunger}(2012)}]{popescu_extracting_2012}%
  \BibitemOpen
  \bibfield  {author} {\bibinfo {author} {\bibfnamefont {V.}~\bibnamefont
  {Popescu}}\ and\ \bibinfo {author} {\bibfnamefont {A.}~\bibnamefont
  {Zunger}},\ }\bibfield  {title} {\bibinfo {title} {Extracting {$E$} versus
  {$\vec {\mathbf k}$} effective band structure from supercell calculations on
  alloys and impurities},\ }\href@noop {} {\bibfield  {journal} {\bibinfo
  {journal} {Phys. Rev. B}\ }\textbf {\bibinfo {volume} {85}},\ \bibinfo
  {pages} {085201} (\bibinfo {year} {2012})}\BibitemShut {NoStop}%
\bibitem [{\citenamefont {Dirnberger}\ \emph {et~al.}(2021)\citenamefont
  {Dirnberger}, \citenamefont {Kresse}, \citenamefont {Franchini},\ and\
  \citenamefont {Reticcioli}}]{dirnberger_electronic_2021}%
  \BibitemOpen
  \bibfield  {author} {\bibinfo {author} {\bibfnamefont {D.}~\bibnamefont
  {Dirnberger}}, \bibinfo {author} {\bibfnamefont {G.}~\bibnamefont {Kresse}},
  \bibinfo {author} {\bibfnamefont {C.}~\bibnamefont {Franchini}},\ and\
  \bibinfo {author} {\bibfnamefont {M.}~\bibnamefont {Reticcioli}},\ }\bibfield
   {title} {\bibinfo {title} {Electronic {State} {Unfolding} for {Plane}
  {Waves}: {Energy} {Bands}, {Fermi} {Surfaces}, and {Spectral} {Functions}},\
  }\href {https://doi.org/10.1021/acs.jpcc.1c02318} {\bibfield  {journal}
  {\bibinfo  {journal} {J. Phys. Chem. C}\ }\textbf {\bibinfo {volume} {125}},\
  \bibinfo {pages} {12921} (\bibinfo {year} {2021})}\BibitemShut {NoStop}%
\bibitem [{Note3()}]{Note3}%
  \BibitemOpen
  \bibinfo {note} {In a practical device, one would expect the MoS2 monolayer
  to be partially constrained by the substrate, gate dielectric, contacts,
  etc., that preclude complete structural relaxation, the unrelaxed and
  (free-standing) fully relaxed cases serving as limiting cases.}\BibitemShut
  {Stop}%
\bibitem [{\citenamefont {Zhu}\ \emph {et~al.}(2011)\citenamefont {Zhu},
  \citenamefont {Cheng},\ and\ \citenamefont
  {Schwingenschlögl}}]{zhu_giant_2011}%
  \BibitemOpen
  \bibfield  {author} {\bibinfo {author} {\bibfnamefont {Z.~Y.}\ \bibnamefont
  {Zhu}}, \bibinfo {author} {\bibfnamefont {Y.~C.}\ \bibnamefont {Cheng}},\
  and\ \bibinfo {author} {\bibfnamefont {U.}~\bibnamefont
  {Schwingenschlögl}},\ }\bibfield  {title} {\bibinfo {title} {Giant
  spin-orbit-induced spin splitting in two-dimensional transition-metal
  dichalcogenide semiconductors},\ }\href
  {https://doi.org/10.1103/PhysRevB.84.153402} {\bibfield  {journal} {\bibinfo
  {journal} {Phys. Rev. B}\ }\textbf {\bibinfo {volume} {84}},\ \bibinfo
  {pages} {153402} (\bibinfo {year} {2011})}\BibitemShut {NoStop}%
\bibitem [{\citenamefont {Ramasubramaniam}(2012)}]{ramasubramaniam_large_2012}%
  \BibitemOpen
  \bibfield  {author} {\bibinfo {author} {\bibfnamefont {A.}~\bibnamefont
  {Ramasubramaniam}},\ }\bibfield  {title} {\bibinfo {title} {Large excitonic
  effects in monolayers of molybdenum and tungsten dichalcogenides},\ }\href
  {https://doi.org/10.1103/PhysRevB.86.115409} {\bibfield  {journal} {\bibinfo
  {journal} {Phys. Rev. B}\ }\textbf {\bibinfo {volume} {86}},\ \bibinfo
  {pages} {115409} (\bibinfo {year} {2012})}\BibitemShut {NoStop}%
\bibitem [{\citenamefont {Kormányos}\ \emph {et~al.}(2013)\citenamefont
  {Kormányos}, \citenamefont {Zólyomi}, \citenamefont {Drummond},
  \citenamefont {Rakyta}, \citenamefont {Burkard},\ and\ \citenamefont
  {Fal'ko}}]{kormanyos_monolayer_2013}%
  \BibitemOpen
  \bibfield  {author} {\bibinfo {author} {\bibfnamefont {A.}~\bibnamefont
  {Kormányos}}, \bibinfo {author} {\bibfnamefont {V.}~\bibnamefont
  {Zólyomi}}, \bibinfo {author} {\bibfnamefont {N.~D.}\ \bibnamefont
  {Drummond}}, \bibinfo {author} {\bibfnamefont {P.}~\bibnamefont {Rakyta}},
  \bibinfo {author} {\bibfnamefont {G.}~\bibnamefont {Burkard}},\ and\ \bibinfo
  {author} {\bibfnamefont {V.~I.}\ \bibnamefont {Fal'ko}},\ }\bibfield  {title}
  {\bibinfo {title} {Monolayer {MoS$_2$}: {Trigonal} warping, the {$\Gamma$}
  valley, and spin-orbit coupling effects},\ }\href
  {https://doi.org/10.1103/PhysRevB.88.045416} {\bibfield  {journal} {\bibinfo
  {journal} {Phys. Rev. B}\ }\textbf {\bibinfo {volume} {88}},\ \bibinfo
  {pages} {045416} (\bibinfo {year} {2013})}\BibitemShut {NoStop}%
\bibitem [{\citenamefont {Marinov}\ \emph {et~al.}(2017)\citenamefont
  {Marinov}, \citenamefont {Avsar}, \citenamefont {Watanabe}, \citenamefont
  {Taniguchi},\ and\ \citenamefont {Kis}}]{marinov_resolving_2017}%
  \BibitemOpen
  \bibfield  {author} {\bibinfo {author} {\bibfnamefont {K.}~\bibnamefont
  {Marinov}}, \bibinfo {author} {\bibfnamefont {A.}~\bibnamefont {Avsar}},
  \bibinfo {author} {\bibfnamefont {K.}~\bibnamefont {Watanabe}}, \bibinfo
  {author} {\bibfnamefont {T.}~\bibnamefont {Taniguchi}},\ and\ \bibinfo
  {author} {\bibfnamefont {A.}~\bibnamefont {Kis}},\ }\bibfield  {title}
  {\bibinfo {title} {Resolving the spin splitting in the conduction band of
  monolayer {MoS$_2$}},\ }\href {https://doi.org/10.1038/s41467-017-02047-5}
  {\bibfield  {journal} {\bibinfo  {journal} {Nat Commun}\ }\textbf {\bibinfo
  {volume} {8}},\ \bibinfo {pages} {1938} (\bibinfo {year} {2017})}\BibitemShut
  {NoStop}%
\bibitem [{\citenamefont {Cui}\ \emph {et~al.}(2018)\citenamefont {Cui},
  \citenamefont {Xue}, \citenamefont {Hu},\ and\ \citenamefont
  {Li}}]{cui_two-dimensional_2018}%
  \BibitemOpen
  \bibfield  {author} {\bibinfo {author} {\bibfnamefont {C.}~\bibnamefont
  {Cui}}, \bibinfo {author} {\bibfnamefont {F.}~\bibnamefont {Xue}}, \bibinfo
  {author} {\bibfnamefont {W.-J.}\ \bibnamefont {Hu}},\ and\ \bibinfo {author}
  {\bibfnamefont {L.-J.}\ \bibnamefont {Li}},\ }\bibfield  {title} {\bibinfo
  {title} {Two-dimensional materials with piezoelectric and ferroelectric
  functionalities},\ }\href {https://doi.org/10.1038/s41699-018-0063-5}
  {\bibfield  {journal} {\bibinfo  {journal} {npj 2D Mater Appl}\ }\textbf
  {\bibinfo {volume} {2}},\ \bibinfo {pages} {1} (\bibinfo {year}
  {2018})}\BibitemShut {NoStop}%
\bibitem [{\citenamefont {Kresse}\ and\ \citenamefont
  {Furthmüller}(1996{\natexlab{a}})}]{kresse_efficient_1996}%
  \BibitemOpen
  \bibfield  {author} {\bibinfo {author} {\bibfnamefont {G.}~\bibnamefont
  {Kresse}}\ and\ \bibinfo {author} {\bibfnamefont {J.}~\bibnamefont
  {Furthmüller}},\ }\bibfield  {title} {\bibinfo {title} {Efficient iterative
  schemes for ab initio total-energy calculations using a plane-wave basis
  set},\ }\href {https://doi.org/10.1103/PhysRevB.54.11169} {\bibfield
  {journal} {\bibinfo  {journal} {Phys. Rev. B}\ }\textbf {\bibinfo {volume}
  {54}},\ \bibinfo {pages} {11169} (\bibinfo {year}
  {1996}{\natexlab{a}})}\BibitemShut {NoStop}%
\bibitem [{\citenamefont {Kresse}\ and\ \citenamefont
  {Furthmüller}(1996{\natexlab{b}})}]{kresse_efficiency_1996}%
  \BibitemOpen
  \bibfield  {author} {\bibinfo {author} {\bibfnamefont {G.}~\bibnamefont
  {Kresse}}\ and\ \bibinfo {author} {\bibfnamefont {J.}~\bibnamefont
  {Furthmüller}},\ }\bibfield  {title} {\bibinfo {title} {Efficiency of
  ab-initio total energy calculations for metals and semiconductors using a
  plane-wave basis set},\ }\href {https://doi.org/10.1016/0927-0256(96)00008-0}
  {\bibfield  {journal} {\bibinfo  {journal} {Comput. Mater. Sci.}\
  }\textbf {\bibinfo {volume} {6}},\ \bibinfo {pages} {15} (\bibinfo {year}
  {1996}{\natexlab{b}})}\BibitemShut {NoStop}%
\bibitem [{\citenamefont {Blöchl}(1994)}]{blochl_projector_1994}%
  \BibitemOpen
  \bibfield  {author} {\bibinfo {author} {\bibfnamefont {P.~E.}\ \bibnamefont
  {Blöchl}},\ }\bibfield  {title} {\bibinfo {title} {Projector augmented-wave
  method},\ }\href {https://doi.org/10.1103/PhysRevB.50.17953} {\bibfield
  {journal} {\bibinfo  {journal} {Phys. Rev. B}\ }\textbf {\bibinfo {volume}
  {50}},\ \bibinfo {pages} {17953} (\bibinfo {year} {1994})}\BibitemShut
  {NoStop}%
\bibitem [{\citenamefont {Kresse}\ and\ \citenamefont
  {Joubert}(1999)}]{kresse_ultrasoft_1999}%
  \BibitemOpen
  \bibfield  {author} {\bibinfo {author} {\bibfnamefont {G.}~\bibnamefont
  {Kresse}}\ and\ \bibinfo {author} {\bibfnamefont {D.}~\bibnamefont
  {Joubert}},\ }\bibfield  {title} {\bibinfo {title} {From ultrasoft
  pseudopotentials to the projector augmented-wave method},\ }\href
  {https://doi.org/10.1103/PhysRevB.59.1758} {\bibfield  {journal} {\bibinfo
  {journal} {Phys. Rev. B}\ }\textbf {\bibinfo {volume} {59}},\ \bibinfo
  {pages} {1758} (\bibinfo {year} {1999})}\BibitemShut {NoStop}%
\bibitem [{\citenamefont {Perdew}\ \emph {et~al.}(1996)\citenamefont {Perdew},
  \citenamefont {Burke},\ and\ \citenamefont
  {Ernzerhof}}]{perdew_generalized_1996}%
  \BibitemOpen
  \bibfield  {author} {\bibinfo {author} {\bibfnamefont {J.~P.}\ \bibnamefont
  {Perdew}}, \bibinfo {author} {\bibfnamefont {K.}~\bibnamefont {Burke}},\ and\
  \bibinfo {author} {\bibfnamefont {M.}~\bibnamefont {Ernzerhof}},\ }\bibfield
  {title} {\bibinfo {title} {Generalized {Gradient} {Approximation} {Made}
  {Simple}},\ }\href {https://doi.org/10.1103/PhysRevLett.77.3865} {\bibfield
  {journal} {\bibinfo  {journal} {Phys. Rev. Lett.}\ }\textbf {\bibinfo
  {volume} {77}},\ \bibinfo {pages} {3865} (\bibinfo {year}
  {1996})}\BibitemShut {NoStop}%
\bibitem [{\citenamefont {Ramasubramaniam}\ \emph {et~al.}(2019)\citenamefont
  {Ramasubramaniam}, \citenamefont {Wing},\ and\ \citenamefont
  {Kronik}}]{ramasubramaniam_transferable_2019}%
  \BibitemOpen
  \bibfield  {author} {\bibinfo {author} {\bibfnamefont {A.}~\bibnamefont
  {Ramasubramaniam}}, \bibinfo {author} {\bibfnamefont {D.}~\bibnamefont
  {Wing}},\ and\ \bibinfo {author} {\bibfnamefont {L.}~\bibnamefont {Kronik}},\
  }\bibfield  {title} {\bibinfo {title} {Transferable screened range-separated
  hybrids for layered materials: {The} cases of {MoS$_2$} and h-{BN}},\ }\href
  {https://doi.org/10.1103/PhysRevMaterials.3.084007} {\bibfield  {journal}
  {\bibinfo  {journal} {Phys. Rev. Mater.}\ }\textbf {\bibinfo {volume} {3}},\
  \bibinfo {pages} {084007} (\bibinfo {year} {2019})}\BibitemShut {NoStop}%
\bibitem [{\citenamefont {Krukau}\ \emph {et~al.}(2006)\citenamefont {Krukau},
  \citenamefont {Vydrov}, \citenamefont {Izmaylov},\ and\ \citenamefont
  {Scuseria}}]{krukau_influence_2006}%
  \BibitemOpen
  \bibfield  {author} {\bibinfo {author} {\bibfnamefont {A.~V.}\ \bibnamefont
  {Krukau}}, \bibinfo {author} {\bibfnamefont {O.~A.}\ \bibnamefont {Vydrov}},
  \bibinfo {author} {\bibfnamefont {A.~F.}\ \bibnamefont {Izmaylov}},\ and\
  \bibinfo {author} {\bibfnamefont {G.~E.}\ \bibnamefont {Scuseria}},\
  }\bibfield  {title} {\bibinfo {title} {Influence of the exchange screening
  parameter on the performance of screened hybrid functionals},\ }\href
  {https://doi.org/10.1063/1.2404663} {\bibfield  {journal} {\bibinfo
  {journal} {J. Chem. Phys.}\ }\textbf {\bibinfo {volume}
  {125}},\ \bibinfo {pages} {224106} (\bibinfo {year} {2006})}%
  \bibitem [{\citenamefont {Shi}\ \emph {et~al.}(2018)\citenamefont {Shi},
  \citenamefont {Guo}, \citenamefont {Zhang},\ and\ \citenamefont
  {Guo}}]{shi_flexo_2018}%
  \BibitemOpen
  \bibfield  {author} {\bibinfo {author} {\bibfnamefont {W.}\ \bibnamefont
  {Shi}}, \bibinfo {author} {\bibfnamefont {Y.}\ \bibnamefont {Guo}},
  \bibinfo {author} {\bibfnamefont {Z.}\ \bibnamefont {Zhang}},\ and\
  \bibinfo {author} {\bibfnamefont {W.}\ \bibnamefont {Guo}},\
  }\bibfield  {title} {\bibinfo {title} {Flexoelectricity in Monolayer Transition Metal 
  Dichalcogenides},\ }\href {https://doi.org/10.1021/acs.jpclett.8b03325} 
  {\bibfield  {journal} {\bibinfo {journal} {J. Phys. Chem. Lett.}\ }\textbf {\bibinfo {volume}
  {9}},\ \bibinfo {pages} {6841} (\bibinfo {year} {2018})}%
  \bibitem [{\citenamefont {Yang}\ \emph {et~al.}(2022)\citenamefont {Yang},
  \citenamefont {Hirsinger}, \ and\ \citenamefont
  {Devel}}]{devel_flexo_2022}%
   \BibitemOpen
  \bibfield  {author} {\bibinfo {author} {\bibfnamefont {Y.}\ \bibnamefont
  {Yang}}, \bibinfo {author} {\bibfnamefont {L.}\ \bibnamefont {Hirsinger}},
  \ and\ \bibinfo {author} {\bibfnamefont {M.}\ \bibnamefont {Devel}},\
  }\bibfield  {title} {\bibinfo {title} {Computation of flexoelectric coefficients of a MoS$_2$ 
  monolayer with a model of self-consistently distributed effective charges and dipoles},\ }
  \href {https://doi.org/10.1063/5.0088972} 
  {\bibfield  {journal} {\bibinfo {journal} {J. Chem. Phys.}\ }\textbf {\bibinfo {volume}
  {156}},\ \bibinfo {pages} {174104} (\bibinfo {year} {2022})}\BibitemShut
  {NoStop}%
\end{thebibliography}

%

\end{document}